\documentclass[10pt,twocolumn]{article}
\usepackage{times}

\usepackage{url}
\usepackage{xspace}
\usepackage{etoolbox}
\usepackage{enumitem}
\usepackage{comment}
\usepackage{booktabs}
\usepackage{graphicx}
\usepackage{makecell}
\usepackage{multirow}
\usepackage{subcaption}
\usepackage{tikz}
\usepackage{caption}
\usepackage{flushend}
\usepackage[shortcuts]{extdash}

\mathchardef\mhyphen="2D %

\newcommand{\myparagraph}[1]{\noindent\textbf{#1. }}
\newcommand{\eg}{\emph{e.g.}\xspace}

\newcommand{\ie}{\emph{i.e.}\xspace}

\newcommand{\faast}{\texttt{Faa\$T}\xspace}

\DeclareRobustCommand\circled[1]{\tikz[baseline=(char.base)]{
            \node[shape=circle,draw,inner sep=1pt] (char) {#1};}}

\begin{document}
\title{
\faast: A Transparent Auto-Scaling Cache for Serverless Applications
}

\author{
Francisco Romero$^{\S}$\thanks{Romero is affiliated with Stanford University,
  but was at Microsoft Research during this work.} \hspace{.01in}
Gohar Irfan Chaudhry$^\dag$
\'I\~{n}igo Goiri$^\dag$
Pragna Gopa$^\ddag$
Paul Batum$^\ddag$ \\
Neeraja J. Yadwadkar$^\S$
Rodrigo Fonseca$^\dag$
Christos Kozyrakis$^\S$
Ricardo Bianchini$^\dag$
\\
\small{
$^\dag$Microsoft Research
\hspace{.3in}
$^\S$Stanford University
\hspace{.3in}
$^\ddag$Microsoft Azure
}
}
\date{\vspace*{-.25in}}

\maketitle

\begin{abstract}
Function-as-a-Service (FaaS) has become an increasingly popular way
for users to deploy their applications without the burden of managing
the underlying infrastructure.  However, existing FaaS platforms 
rely on remote storage to maintain state, limiting the
set of applications that can be run efficiently.  Recent caching work
for FaaS platforms has tried to address this problem, but has fallen
short: it disregards the widely different characteristics of FaaS
applications, does not scale the cache based on data access
patterns, or requires changes to applications.  To address these
limitations, we present \faast, a transparent auto-scaling distributed
cache for serverless applications.  Each application gets its own
\faast cache.  After a function executes and the application becomes
inactive, the cache is unloaded from memory with the application.
Upon reloading for the next invocation, \faast pre-warms the
cache with objects likely to be accessed.  In addition to
traditional compute-based scaling, \faast scales based on working set
and object sizes to manage cache space and I/O bandwidth.  We motivate
our design with a comprehensive study of data access patterns in a
large-scale commercial FaaS provider.  We implement \faast for the
provider's production FaaS platform.  
Our experiments show that \faast can improve performance by up to 92\% (57\%
on average) for challenging applications, and reduce cost for most users compared to state-of-the-art caching systems, \ie the cost of having to stand up additional serverful resources.
\end{abstract}

\section{Introduction}

\myparagraph{Motivation}
Function-as-a-Service (FaaS) is an increasingly popular way of
deploying applications to the cloud.  With FaaS, users deploy their code
as stateless functions and need not worry about creating, configuring, or
managing resources explicitly.  FaaS shifts these responsibilities to
the FaaS provider (\eg, AWS Lambda, Azure Functions, Google Cloud
Functions), which charges users per resource usage during their
function invocations.  FaaS providers build their platforms by renting
and managing resources (VMs or bare-metal containers) in public clouds
(\eg, AWS, Azure, Google Cloud
Platform).  To control their costs, these providers proactively unload
a function from memory if it has not been invoked for a while (\eg,
after 7 minutes of inactivity in AWS Lambda \cite{mikhailcoldstart}).

Due to FaaS's stateless nature, a function invocation is not guaranteed
to have access to state created by previous invocations.  Thus, any state
that might be needed later must be persisted in remote
storage.
This also applies to applications with multiple stages
(often expressed as a pipeline or a directed acyclic graph of functions),
with intermediate results passed between invocations. Since existing
FaaS platforms typically do not allow functions to communicate directly,
functions must also write these results to remote storage.

The remote storage can be object-based (\eg, Amazon S3~\cite{s3}, Azure Blob
Storage~\cite{blobstorage}), queues~\cite{queuetriggers}, or in-memory
storage clusters (\eg,
Redis~\cite{redis}, InfiniCache~\cite{wang2020infinicache}, Pocket~\cite{klimovic2018pocket}).
Regardless of type, remote storage incurs higher latency and lower
bandwidth than accessing local memory~\cite{klimovic2018pocket, TwoStepsBack19}.
When users have to provision
in-memory storage clusters, it introduces management
overhead and costs. %

Given these limitations, local in-memory caching emerges as a natural
solution for both speeding up access to remote data and enabling
faster data sharing between functions.  Prior works have considered
both local and remote in-memory caching for
FaaS~\cite{klimovic2018pocket, pu2019locus, sreekanti2020cloudburst,
wang2020infinicache}
but, we argue, have come up short in multiple ways.

First, they implement a single cache for multiple or all applications.
This disregards the widely different characteristics of FaaS applications.
For example, Shahrad {\em et al.}~\cite{shahrad2020serverless} have
shown that 45\% of applications are invoked less frequently than
once per hour on average.
Caching data for rarely-invoked applications at all times is wasteful.
However, not caching data for these applications will likely produce poor performance.
Furthermore, a shared cache requires complex communication and synchronization primitives for the data of thousands of applications.

Second, in prior approaches, the cache is either fixed in size (\eg,
~\cite{pu2019locus, wang2020infinicache}) or scales only according to the
computational load (\eg,~\cite{sreekanti2020cloudburst}).  These
approaches work well when data
access patterns are stable and working sets are smaller than
the available cache space.  When this
is not the case, scaling the cache based on data access patterns
would be beneficial.
Moreover, prior works have not considered scaling as a way to
mitigate the impact of accessing large data objects; these objects
can take long to access as remote storage I/O bandwidth is often limited by the
underlying VM/container or contention across co-located
applications~\cite{TwoStepsBack19}.
Emerging FaaS applications, such as ML inference pipelines and notebooks,
would benefit from scaling out for increased cache space and increased I/O
bandwidth to remote storage.

Third, prior caches are not entirely transparent to users, either
because users need to provision them explicitly
(\eg,~\cite{pu2019locus, klimovic2018pocket}) or because they
provide a separate API for users to access the cache
(\eg,~\cite{shillaker2020faasm, sreekanti2020cloudburst}).  FaaS users
do not want to think of data locality or manage caches (or any other
resources) explicitly.  In fact, a key reason for the popularity of
FaaS is exactly that users are relieved from such tasks.

\myparagraph{Our work} Fundamentally, the problem with prior
approaches is that {\it caching layers for serverless systems have never been truly
serverless}, \ie~tied to applications, auto-scaling, and transparent.
Thus, we propose \faast, an in-memory caching layer with these
characteristics.

Each application is loaded into memory with its own local \faast cache.
This cache manages the data accessed by the application transparently
as it runs.  When the application is unloaded from memory, its \faast is
also unloaded.  This approach obviates the need for a remote
in-memory cache and may reduce the overall traffic to remote storage,
both of which reduce costs for users.  It also means that the cache space
required by rarely-invoked applications is proactively removed from
memory when not needed, just as the application itself, which reduces
costs for the FaaS provider.  Moreover, it enables different cache
replacement and persistence policies per application, and pre-fetching
of the most popular data when (re-)loading each application.  The latter
feature can be very effective when combined with automatic pre-warming of
applications, which can be done accurately and completely off the
critical invocation path~\cite{shahrad2020serverless}.

As in other systems, an application is auto-scaled based on the number
of invocations it is receiving.  Scaling out loads a new application
``instance'' (\ie, a copy of its code) into memory, whereas scaling in
unloads an instance.  We refer to this as compute-based scaling.
However, to match each application's data access and reuse patterns, \faast
automatically scales out the number of instances to (a)
increase the fetch bandwidth from remote storage for large objects,
and (b) increase the overall cache size for frequently-accessed remote
data.  Scale-in occurs as the need for space and/or bandwidth subsides.
With multiple active instances, \faast forms a cooperative distributed
cache, \ie a data access can hit locally, miss locally but hit a
remote cache, or miss both locally and remotely.  By
default, \faast offers strong data consistency,
but each application can {\em optionally} select its own
consistency and policies for scaling and eviction.

A key aspect of FaaS is that users are only charged for resources they
use.  As we tie \faast to applications, we expect that FaaS providers
would charge only for cache accesses and space consumed by objects
that were actually accessed.

\myparagraph{Implementation and results} We motivate \faast with the
first comprehensive study of real FaaS applications from the perspective of data
access and reuse.  We use data collected for 2 weeks from the production workload
of a large public FaaS provider.  We show, for example, that many
infrequently invoked applications exhibit good temporal locality (the
same data is accessed across relatively rare invocations), whereas
spatial locality in large objects is high (if any byte from such an
object is accessed, the rest of it should be pre-fetched).

We implement \faast in a public production FaaS offering.  To show
that it enables new applications that are not efficiently supported by
current FaaS platforms, we implement an ML inference pipeline and a
Jupyter notebooks~\cite{jupyter} server stack that runs unmodified
notebooks in a serverless environment.

Our experiments with these applications evaluate \faast's caching and
scaling policies.  Our results show that \faast can improve their performance
by up to 92\% (57\% on average), and reduce costs for most users compared to state-of-the-art FaaS caching systems, \ie, the cost incurred by the provisioning of additional serverful resources.

\myparagraph{Contributions} In summary, our main contributions are:
\begin{itemize}[wide,labelwidth=!,labelindent=0pt,topsep=0pt,itemsep=-1ex,partopsep=1ex,parsep=1ex]
  \item We characterize the data access patterns of the production workload of a large FaaS provider.
  \item We design and implement \faast, a transparent auto-scaling cache for FaaS applications.
  \item We propose scaling policies for \faast to increase instance bandwidth and overall cache size based on data access patterns and object sizes.
  \item We show that \faast broadens the scope of applications that can run on FaaS with near-native performance, including ML inference pipelines and Jupyter notebooks.
\end{itemize}

\begin{figure*}[t]
  \centering
  \minipage[t]{0.32\linewidth}
    \includegraphics[width=1\textwidth]{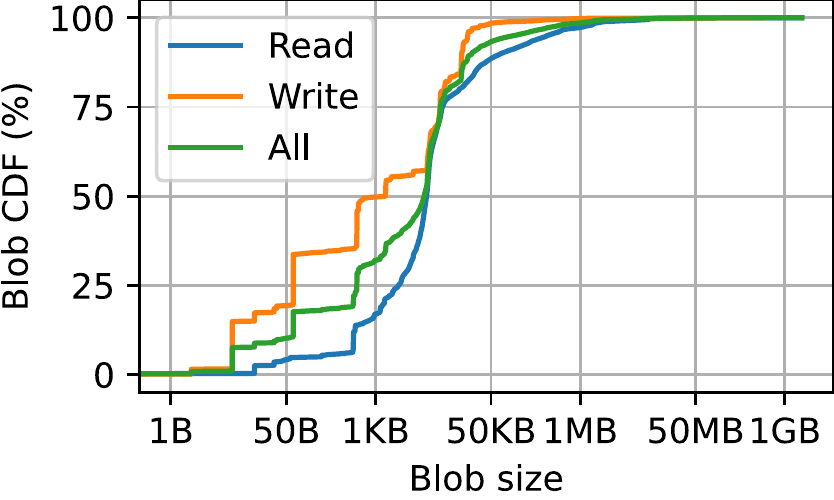}
    \caption{
      Distribution of the size of accessed blobs.
      80\% of accessed blobs are smaller than 12KB.%
    }
    \label{fig:characterization-blobsize}
  \endminipage \hfill
  \minipage[t]{0.32\linewidth}
    \includegraphics[width=1\textwidth]{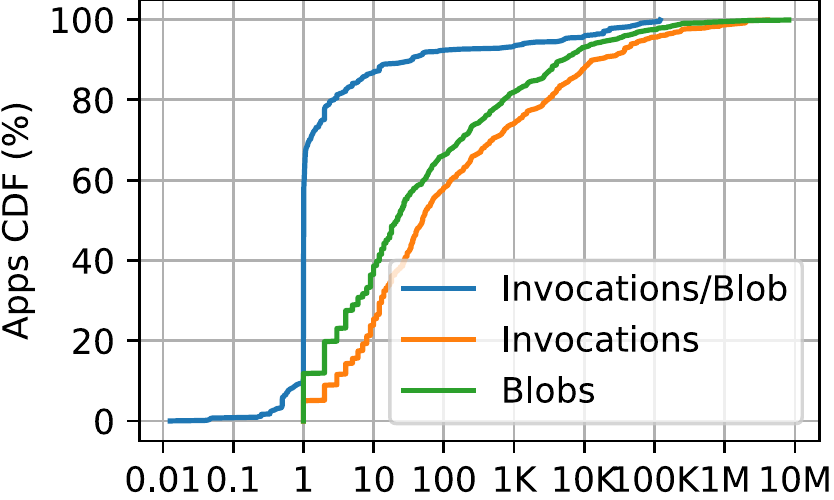}
    \caption{CDF of the number of function invocations per unique
      blob by each application. 
    }
    \label{fig:characterization-accesses-per-app}
  \endminipage \hfill
  \minipage[t]{0.32\linewidth}
    \includegraphics[width=1\textwidth]{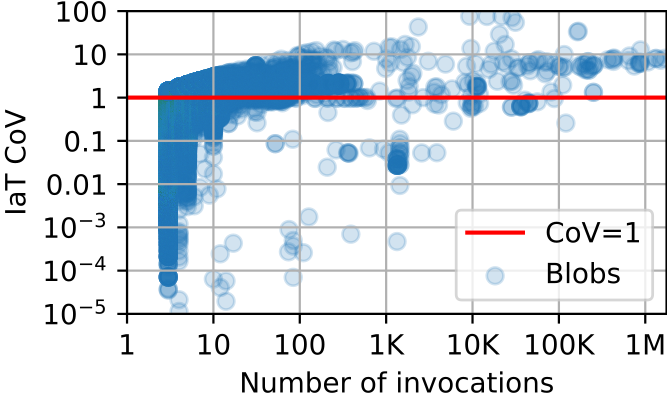}
    \caption{
      CoV of the IaT and number of invocations for each blob accessed more than three times.
      CoV of 1 is Poisson arrival.
    }
    \label{fig:accesses-time}
  \endminipage
  \vspace*{-.2in}
\end{figure*}

\section{Analysis of FaaS Applications and Caching}
\label{sec:apps}

This section characterizes the invocation and data access patterns of
applications running on a production FaaS offering.
An {\em application} is a collection of logically-related functions.
Each instance of an application gets a set of resources (\eg, memory in a VM or
container) that is shared by its functions.
We focus on data accesses, as several prior works have focused on code accesses to optimize cold-start latencies~\cite{agache2020firecracker,oakes2018sock,du2020catalyzer,cadden2020seuss} or reducing the number of cold-starts~\cite{shahrad2020serverless}.
We also discuss the limitations of current FaaS platforms for existing and emerging applications. 

\subsection{Characterizing Current Applications}
\label{sec:character}

We use 14 days of logs (November 23$^{rd}$ to December
7$^{th}$ 2020) from a large-scale FaaS provider, across 28 geographical regions.
We analyze a sample of the
applications that access remote blob storage over HTTPS.  
This log includes 855
applications from 509 users and 33.1 million invocations with 44.3 million data accesses. 77.3\% of 
accesses are reads and the rest are writes. The
applications are written in multiple programming languages,
including C\#, Node.js, and Python.

\myparagraph{Data size}
The log includes accesses to 20.3 million different objects with a total size of 1.9 TB.
Figure~\ref{fig:characterization-blobsize} shows the distribution of the size of blobs accessed.
80\% of blobs are smaller than 12KB and more than 25\% are smaller than 600 bytes.
However, there are also many large blobs; a few as large as 1.8GB.
The objects read tend to be larger than the ones
written.
While the aggregate
bandwidth to backend storage is usually high~\cite{vaishaal20numpywren}, the
prevalence of small objects exacerbates the impact of storage latency.

\myparagraph{Data accesses and reuse}
Figure~\ref{fig:characterization-accesses-per-app} shows the CDF
of the ratio of the number of invocations an application made to the
number of unique blobs accessed. 
Most applications access a single, different blob per invocation (invocation/blob=1).
Roughly 11.0\% of applications access more than one blob per invocation (invocation/blob<1).
More than 32.0\% of the applications access the same blob in more than
one invocation and 7.7\% in more than 100 invocations.
One application accesses the same blob in more than 10,000
invocations.

Around 11.8\% of the applications access the same blob across all invocations,
66.1\% access less than 100 different blobs,
93.0\% access less than 10,000 different blobs,
and one accesses more than 8 million different blobs.
Even though there are 44.3 million accesses, only 20.3 million are first accesses.
Overall, the applications accessed 2.6 TB of data while the corpus of unique data is 1.9 TB.
\emph{If we were able to cache the already accessed data, we could save up to 27.0\% of traffic and 54.3\% of the accesses to remote storage.}

Data sharing across applications and users (not shown) is extremely uncommon.
99.7\% of the blobs are not shared across applications,
only 0.02\% of blobs
are shared across regions,
and only 16 blobs across users.

\myparagraph{Temporal access pattern}
Figure~\ref{fig:accesses-time} shows the temporal access patterns for each blob accessed by an application.
The X-axis is the number of function invocations
that read/wrote the blob and the Y-axis shows the coefficient of variation (CoV) for the inter-arrival time (IaT) of those invocations.
Each point represents a blob with more than 3 accesses (there is no IaT CoV otherwise).
A CoV of 1 suggests Poisson arrivals, values close to 0 indicate a periodic arrival, and values larger than 1 indicate greater burstiness than Poisson.
Clearly, accesses to a large percentage of blobs are very bursty.

\begin{table*}[t]
    \resizebox{\linewidth}{!}{%
    \centering
    \begin{tabular}{c|cccccccc}
        \toprule[1pt]\midrule[0.3pt]
        Data size & Large & Large & Large & Large & Small & Small & Small & Small \\ \midrule
        App. invoc. freq. & Frequent & Rare & Frequent & Rare & Frequent & Rare & Frequent & Rare \\ \midrule
        Data reuse & Low-med & Med-high & Med-high & Med-high & Low & Med & Low-med & Low \\ \midrule
        \makecell{Example \\ application} & \makecell{Video \\ streaming} & \makecell{Jupyter \\ notebook} & \makecell{ML inference \\ (pipeline)} & \makecell{Distributed \\ compilation} & \makecell{Log \\ aggregator} & \makecell{Software \\ unit tests} & \makecell{ML inference \\ (single model)} & \makecell{IoT \\ sensors} \\ \bottomrule
    \end{tabular}
    } %
    \caption{
      FaaS application spectrum.
      Applications vary in data size, invocation frequency, and data reuse.
      }
    \label{tab:app-spectrum}
    \vspace*{-.1in}
\end{table*}

\begin{table*}[t]
    \resizebox{\linewidth}{!}{%
    \centering
    \begin{tabular}{lccccccc}
        \toprule[1pt]\midrule[0.3pt]
        Features/Properties & Pocket~\cite{klimovic2018pocket} & InfiniCache~\cite{wang2020infinicache} & Locus~\cite{pu2019locus} & Faasm~\cite{shillaker2020faasm} & Cloudburst~\cite{sreekanti2020cloudburst} & OFC~\cite{mvondo2021ofc} & \faast \\ \midrule[1pt]
        Cache location & {\color[HTML]{FE0000} Storage cluster} & {\color[HTML]{CD9934} Separate function} & {\color[HTML]{FE0000} Redis cluster} & {\color[HTML]{32CB00} Host cache} & {\color[HTML]{32CB00} VM cache} & {\color[HTML]{32CB00} RAMCloud server} & {\color[HTML]{32CB00} Invoked function} \\
        Cached data management & {\color[HTML]{CD9934} Independent of app} & {\color[HTML]{CD9934} Independent of app} & {\color[HTML]{CD9934} Independent of app} & {\color[HTML]{CD9934} Independent of app} & {\color[HTML]{CD9934} Independent of app} & {\color[HTML]{32CB00} Unloaded w/ app} & {\color[HTML]{32CB00} Unloaded w/ app} \\ \midrule[0.3pt]
        App transparency & {\color[HTML]{CD9934} Get/Put} & {\color[HTML]{CD9934} Get/Put} & {\color[HTML]{CD9934} No changes$^*$} & {\color[HTML]{FE0000} Custom API} & {\color[HTML]{FE0000} Custom API} & {\color[HTML]{32CB00} No changes} & {\color[HTML]{32CB00} No changes} \\
        User configuration & {\color[HTML]{CD9934} Hints} & {\color[HTML]{CD9934} Num instances} & {\color[HTML]{CD9934} Redis} & {\color[HTML]{32CB00} None} & {\color[HTML]{32CB00} None} & {\color[HTML]{32CB00} None} & {\color[HTML]{32CB00} None} \\
        Data consistency & {\color[HTML]{FE0000} None} & {\color[HTML]{FE0000} None} & {\color[HTML]{FE0000} None} & {\color[HTML]{32CB00} Supported} & {\color[HTML]{32CB00} Supported} & {\color[HTML]{32CB00} Supported} & {\color[HTML]{32CB00} Supported} \\
        Object pre-warming & {\color[HTML]{FE0000} None} & {\color[HTML]{FE0000} None} & {\color[HTML]{FE0000} None} & {\color[HTML]{FE0000} None} & {\color[HTML]{FE0000} None} & {\color[HTML]{FE0000} None} & {\color[HTML]{32CB00} Automatic} \\ \midrule[0.3pt]
        Compute cache scaling & N/A & N/A & N/A & {\color[HTML]{32CB00} Dynamic} & {\color[HTML]{32CB00} Dynamic} & {\color[HTML]{32CB00} Dynamic} & {\color[HTML]{32CB00} Dynamic} \\
        Cache size scaling & N/A & {\color[HTML]{FE0000} None} & {\color[HTML]{FE0000} None} & {\color[HTML]{FE0000} None} & {\color[HTML]{FE0000} None} & {\color[HTML]{32CB00} Dynamic} & {\color[HTML]{32CB00} Dynamic} \\
        Cache bandwidth scaling & N/A & {\color[HTML]{FE0000} None} & {\color[HTML]{FE0000} None} & {\color[HTML]{FE0000} None} & {\color[HTML]{FE0000} None} & {\color[HTML]{FE0000} None} & {\color[HTML]{32CB00} Dynamic} \\ \bottomrule
    \end{tabular}
    } %
    \caption{
      Existing systems and their properties.
      Green indicates supported/enabled.
      Tan indicates limited support, or limitations in what can be enabled.
      Red indicates not supported/enabled.
      For Locus, $^*$ indicates transparency is limited to applications with a shuffle operation.
      }
    \label{tab:existing-systems}
    \vspace{-.2in}
\end{table*}

\myparagraph{Access performance}
We observe that writes are usually faster than reads, since
writes are buffered and do not require persisting
data across all replicas synchronously.
Reads are slower as we have to wait for the storage layer to deliver
all data.
We also observe that smaller blobs have a lower throughput
(in MB/s) as they cannot amortize the overhead of the initial handshake.

\myparagraph{Diverse invocation patterns}
Prior work has observed that invocation frequency varies greatly: 81\% of applications are invoked at most once a minute on average, whereas 45\% are invoked less than once per hour on average~\cite{shahrad2020serverless}.
Furthermore, less than 20\% of applications are responsible for over 99\% of all invocations.
These findings are consistent with our log where we observe an even more concentrated behavior:
80\% of functions have less than one invocation per minute on average
and less than 15\% of applications account for 99\% of the invocations.
This heterogeneous behavior is a challenge, as caching data for rarely-invoked applications can be wasteful but is necessary, as it affects a large number of users.

\myparagraph{Takeaways and requirements} Our characterization
indicates that many FaaS applications exhibit data reuse: more than
30\% of them access the same data across invocations.  This suggests
that caching can be effective for them.  Moreover, the
characterization shows a wide spectrum of accessed data sizes and
invocation frequencies.  Accessed data sizes span almost 9 orders of
magnitude (from several bytes to GBs), \ie large objects cannot be
overlooked.  Function invocation frequencies also span almost 9 orders
of magnitude, \ie rarely-invoked applications cannot be overlooked.

Table~\ref{tab:app-spectrum} illustrates the spectrum of data sizes,
invocation frequencies, and reuse along with some example
applications.  For instance, distributed compilation of the Chromium
browser requires accesses to hundreds of MBs, but happens only a few
times per day using a framework like gg~\cite{fouladi2019gg}.  Data
reuse across compilations is high since the codebase does not change
fast.  In contrast, serving an IoT sensor involves a small dataset,
rare invocations, and low reuse.

We draw three key requirements for a serverless caching layer.
First, it should ensure that {\em data with good temporal locality
is cached} and reused across invocations (R1).
Second, the caching layer should target \emph{both frequently- and
  rarely-invoked applications} (R2).
It should optimize for data reuse for frequently-invoked applications,
while it should have the ability to pre-warm frequently-accessed
objects for rarely-invoked applications.
Finally, the caching layer should {\em accommodate large objects and
  exploit spatial locality} for them (R3).
These requirements must be satisfied for applications written across various programming languages.

\subsection{Existing Caching Systems Limitations}
\label{sec:existingsys}
Table~\ref{tab:existing-systems} lists the characteristics of several caching systems.

\myparagraph{Caching is managed independently of each application}
Except for OFC, all the systems listed in the 
table include a separate caching or storage
infrastructure that is shared by multiple applications. (Cloudburst and Faasm also have a shared cache on the same hosts/VMs that run the functions.)
Because of
this, either users must manage the extra servers or cache
state is left behind when applications are unloaded from memory. In the
former case, user costs and management overheads are greater, whereas in the
latter the FaaS provider costs are higher. 
Thus, a serverless caching layer \emph{should be tied to each application, so that its code and data can be loaded/unloaded based on the application's invocation pattern} (R4).

\myparagraph{Need system configuration or application changes}
Pocket, InfiniCache, and Locus rely on user configuration to maximize performance and minimize cost.  For example, 
InfiniCache users must statically set the number of data shards and functions
to store them, whereas Locus users must configure a Redis cluster.
Faasm and Cloudburst rely on custom APIs to give applications control over the consistency of their data, and to pass messages between functions.
To retain the simplicity of the FaaS paradigm, the caching layer \emph{should be transparent and not require changes to the application} (R5).

Cloudburst additionally requires its key-value store (Anna \cite{wu2018anna}) to track all objects residing in all caches.
Managing this much metadata can lead to scaling limitations and extra costs.
The caching layer can mitigate these concerns by \emph{minimizing metadata management for each cache instance} (R6).

\myparagraph{Compute-based scaling only}
Finally, while some existing systems do not dynamically scale their caches (Pocket, InfiniCache, Locus), others do so based solely on the amount of offered computational load, \ie number of function invocations (Faasm, Cloudburst).
OFC scales based on the computational load and predicted memory usage of cached objects.
However, it limits its caching to objects smaller than 10MB.
As applications become more complex and data-heavy, data access characteristics like large working sets or large objects will gain in importance.
Thus, the caching layer \emph{should scale compute (as the application's offered load varies), cache size (based on the data reuse pattern), and bandwidth to remote storage (based on the object sizes being accessed)} (R7).

\subsection{Enabling New FaaS Applications}
\label{sec:newapps}
Current FaaS platforms limit the set of applications that can be
efficiently run.  Next, we describe two challenging ones.

\begin{figure}[t]
  \centering
  \includegraphics[width=\columnwidth]{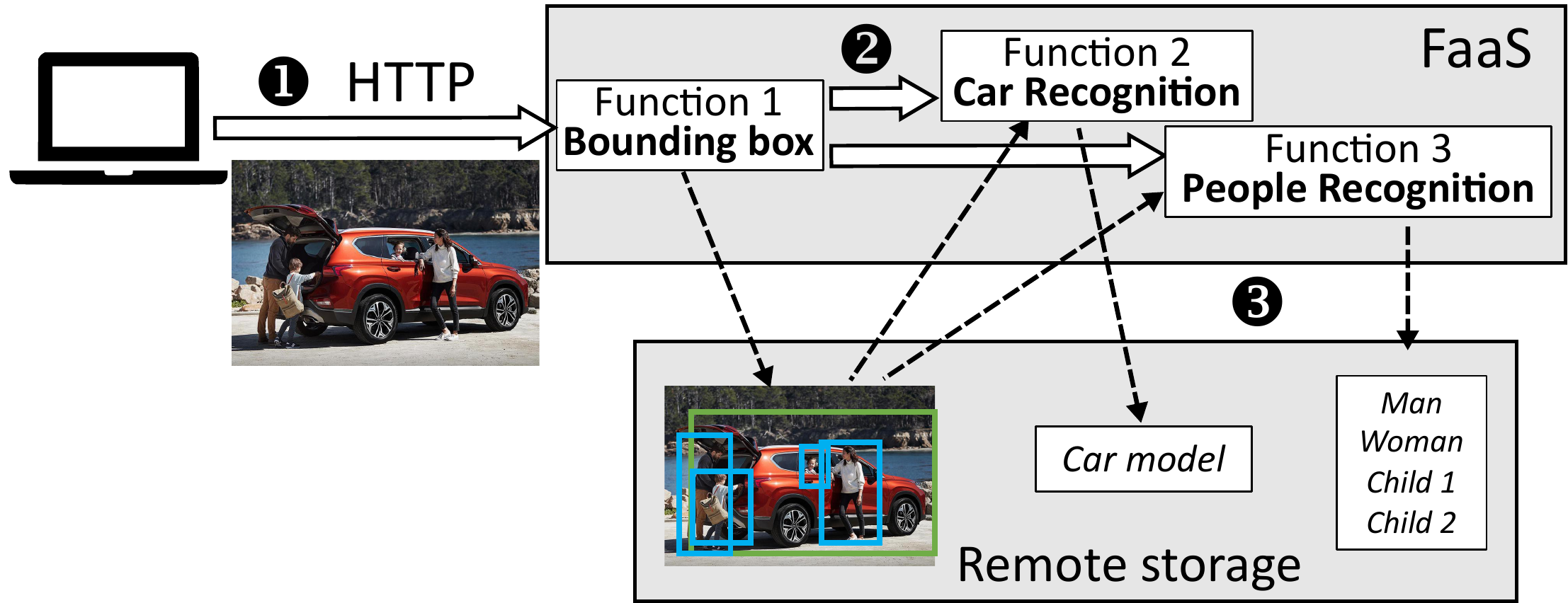}
  \caption{
    An example of ML pipeline executed through FaaS.
    The bounding box model function is invoked using HTTP \circled{1}.
    The labeled bounding boxed image is uploaded to a common data store, and it subsequently triggers the execution of the car and people recognition functions \circled{2}.
    The car and people recognition functions also upload their outputs to the store \circled{3}.
  }
  \label{fig:mlpipeline-diagram}
  \vspace{-.2in}
\end{figure}

\myparagraph{ML inference pipeline}
Many applications across several domains (\eg, health care~\cite{healthcare}, advertisement recommendation~\cite{facebook-recommend}, and retail~\cite{amazongo}) depend on ML inference for classification and other prediction tasks.
ML inference load patterns can vary unpredictably~\cite{infaas_hotos,mlperf}, which makes FaaS on-demand compute and scaling an ideal match for serving inference queries.
However, ML inference applications require low-latency prediction serving ($<$1 s)~\cite{gupta2020deeprecsys,mlperf}.
For example, AMBER Alert~\cite{amber-alerts} responders may use an application to perform car and facial recognition. The application can be deployed on a FaaS platform as a pipeline of ML models (Figure~\ref{fig:mlpipeline-diagram}).
For each input image, an HTTP request is first sent to a bounding box model function to identify and label all present objects \circled{1}.
The labeled image is uploaded to a common data store to trigger the car and people recognition models \circled{2}.
Both recognition functions upload their outputs to the common data store \circled{3}.
Inference pipelines can exhibit different levels of parallelism at each stage, which also makes them good fits for FaaS deployment~\cite{romero2021llama}.
The AMBER Alert pipeline fans out in the second stage, depending on the identified objects.

\begin{figure}[t]
  \centering
  \begin{subfigure}[t]{0.47\linewidth}
    \includegraphics[width=1\columnwidth]{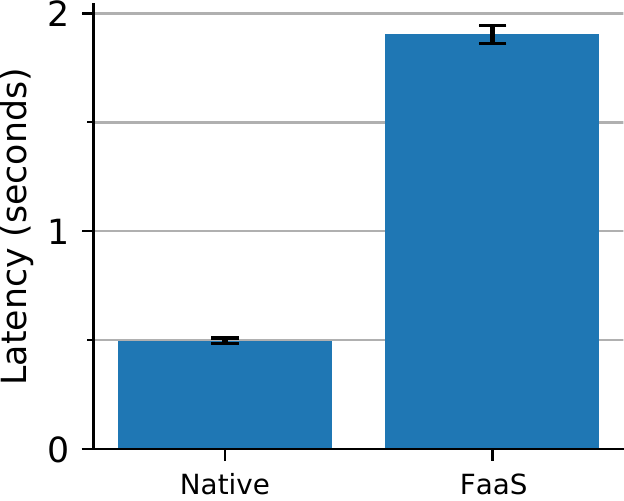}
    \caption{Native VM versus prod FaaS.}
    \label{fig:mlinfmotivall}
  \end{subfigure}
  \begin{subfigure}[t]{0.47\linewidth}
    \includegraphics[width=1\columnwidth]{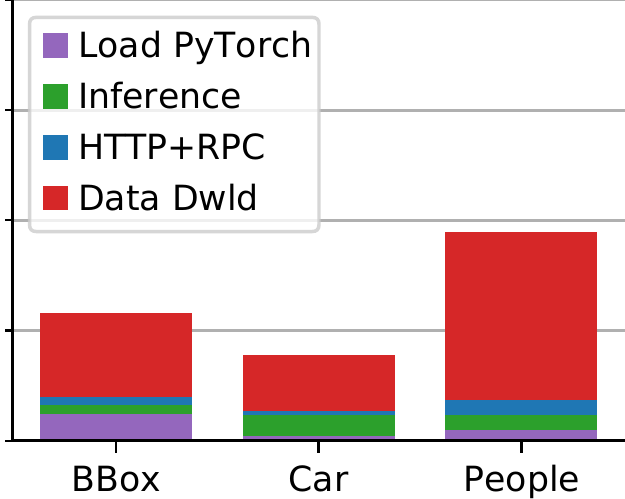}
    \caption{FaaS execution breakdown.}
    \label{fig:mlinfmotivbreakdown}
  \end{subfigure}
  \caption{
    (a) Latency of AMBER Alert pipeline on a native VM versus a production FaaS.
    Native VM does not include the time to load the PyTorch library (700MB).
    It is up to 3.8$\times$ slower to run the pipeline on FaaS.
    (b) Model run times (BBox is a bounding box model).
    Data movement from/to remote storage dominates.
  }
  \label{fig:mlinfmotiv}
  \vspace{-.2in}
\end{figure}

To assess whether FaaS can meet low-latency requirements, we ran the AMBER Alert pipeline natively on a local VM, and in a production FaaS environment with remote storage.
Figure~\ref{fig:mlinfmotivall} shows that it is up to 3.8$\times$ slower running FaaS versus natively, while Figure~\ref{fig:mlinfmotivbreakdown} shows that the main reason is the time to load the models. The inefficiency of the storage layer makes it impossible for the FaaS platform to run this application with sub-second latency.

\myparagraph{Jupyter notebooks}
Jupyter notebooks~\cite{jupyter} are often used for data science tasks such as data processing, ML modeling, and visualizing results~\cite{psallidas2019jupyter}.
They are typically run by defining code in cells and interactively executing each cell. Jupyter notebooks are typically backed by statically-allocated VMs.
Depending on how often a notebook is run, the VMs may sit idle for long periods.
This is expensive for users and wasteful for service providers.
Furthermore, the amount of parallelism and compute needed for \emph{each cell's task} can vary.
Akin to ML inference, this variability makes FaaS a strong fit.

To test its performance, we ported Jupyter to run on a production FaaS platform --- an application we term \emph{JupyterLess}.
Each cell is executed as a function invocation and the state between functions is shared through an intermediate storage layer.
We compare the execution time of summing a single 350MB DataFrame column partitioned into 10 chunks with JupyterLess to running on a native VM.
JupyterLess is 63$\times$ slower than native VM execution as downloading the intermediate state and DataFrame column from remote storage dominates the execution time.
Thus, JupyterLess cannot be run interactively on existing FaaS frameworks.

\section{\faast design}
\label{sec:design}

We design \faast as a transparent auto-scaling cache that meets the requirements we identify in Section~\ref{sec:apps}.
\faast caches objects accessed during a function execution so they can be reused across invocations (R1).
It is built into the FaaS runtime with no external servers or storage layers, so it can be transparently tied to an application (R4, R5) written in any of the various supported languages.
When an application is unloaded from memory, \faast collects metadata about the cache objects, and uses it to pre-warm the cache with frequently accessed objects when the application is re-loaded into memory.
This is especially important for applications that are rarely-invoked (R2).

\faast scales along three dimensions (R7):
(a) based on an application's invocations per second (\emph{compute} scaling), which benefits applications that are frequently-invoked;
(b) based on the data reuse pattern (\emph{cache size} scaling), which is beneficial for applications with large working sets whose objects are continuously evicted;
(c) based on the object size (\emph{bandwidth} scaling), which is
beneficial for applications that access large objects ($\geq$10MB) and are limited by the I/O bandwidth between the application instance and remote storage (R3).
While an application is loaded, \faast efficiently locates objects across instances using consistent hashing without the need for large location metadata (R6).

\subsection{System architecture} \label{sec:arch}
\begin{figure}[t]
  \centering
  \includegraphics[width=.95\columnwidth]{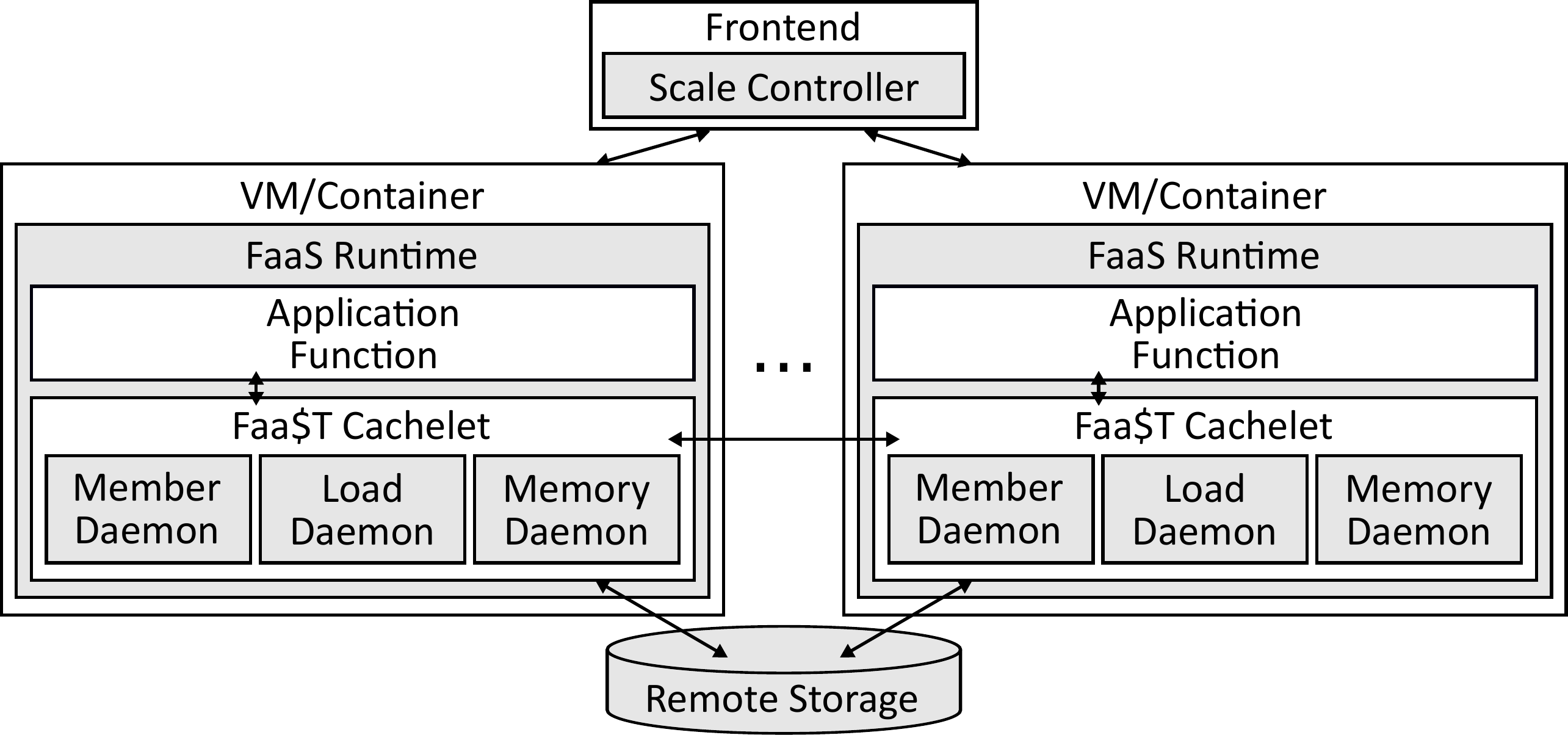}
  \caption{
    \faast's architecture diagram.
  }
  \label{fig:arch-diagram}
  \vspace{-.1in}
\end{figure}

Figure~\ref{fig:arch-diagram} shows the architecture of a FaaS platform with \faast.
Each application instance runs in a VM or container that contains the FaaS runtime and the code for the application functions.
\faast instances, which we refer to as \emph{cachelets}, are a part of the runtime, caching data in memory. 
Each application instance has one corresponding \faast cachelet.
In addition, \faast forms a cooperative distributed cache; an application's \faast cachelets communicate directly to access data when necessary (Section~\ref{sec:access-data}).
Similar to Faasm, \faast maintains a single copy of cached objects, which improves memory efficiency compared to existing systems~\cite{mvondo2021ofc,sreekanti2020cloudburst,wang2020infinicache}.

We designed \faast to be per-application due to the following drawbacks of a shared cache.
First, a shared cache requires complex communication and synchronization primitives for the data of thousands of applications (compared to a maximum of tens of instances for a single application with its own cache).
This makes it difficult to implement per-application management policies (\eg scaling) and provide transparency without custom APIs~\cite{shillaker2020faasm,sreekanti2020cloudburst}, especially given the diversity of application characteristics and requirements (Section~\ref{sec:character}).
Second, a shared cache with traditional eviction policies (\eg LRU) can lead to severe unfairness among applications~\cite{pu2016fairride}.

To find the location of an object, a \faast cachelet interacts with the \emph{Membership Daemon}, which determines the object's ``owner'' based on the current number of cachelets.
The owner is responsible for downloading/uploading the object from/to remote storage.
The \emph{Load Daemon} collects cached object metadata, and uses it to decide what data objects to pre-warm when an application is loaded (Section~\ref{sec:loadunload}).
To prevent interference with an application's heap memory usage, the \emph{Memory Daemon} monitors both function and cachelet memory consumption.
Finally, the \emph{Frontend} load-balances requests across the running application instances, and the \emph{Scale Controller} adds and removes instances based on metrics provided by the FaaS runtime (Section~\ref{sec:scaling}).

\begin{figure}[t]
  \centering
  \includegraphics[width=.65\columnwidth]{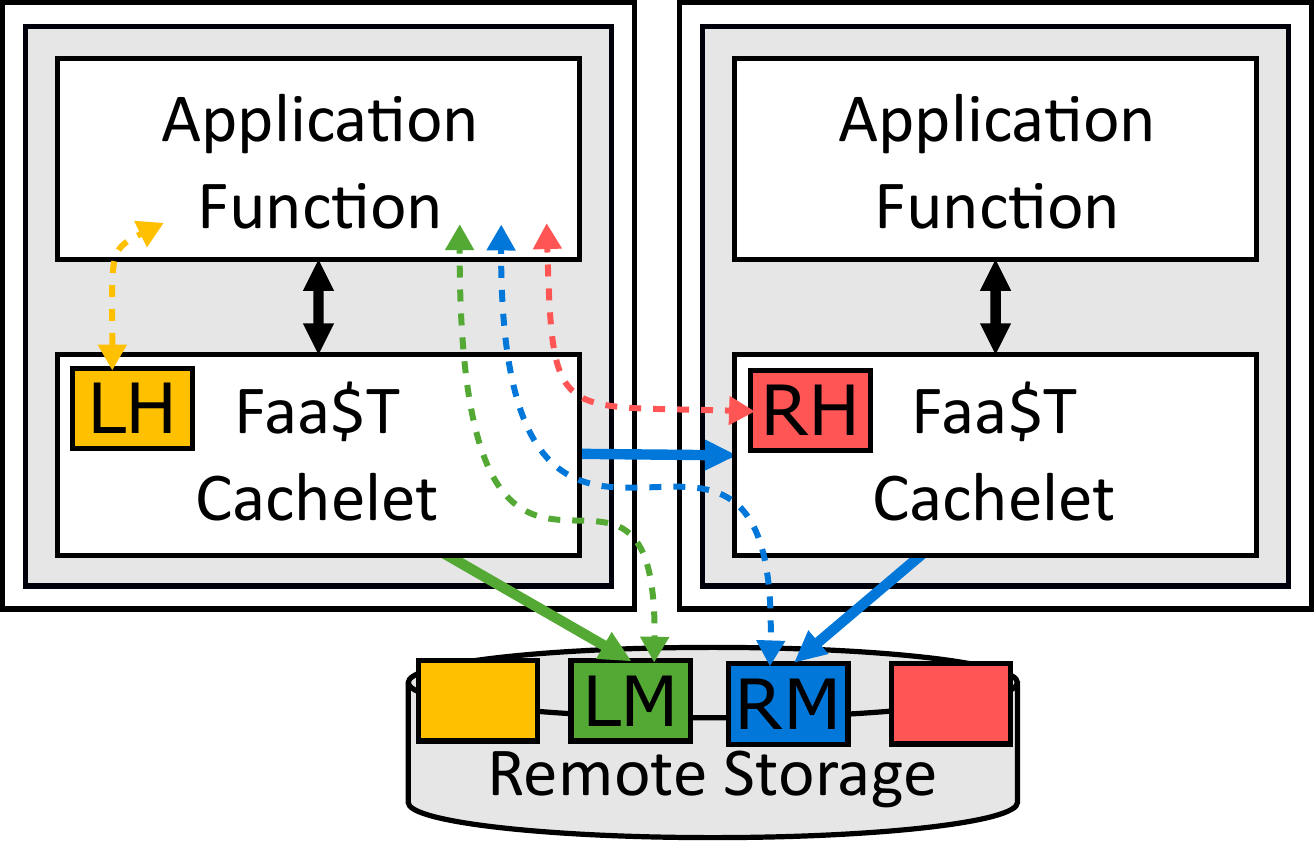}
  \caption{
    Reads in \faast: local hit (LH), remote hit (RH), local miss (LM), remote miss (RM).
    Solid lines indicate communication between the application, \faast instances, and remote storage.
    Dashed lines indicate data movement.
  }
  \label{fig:data-access}
  \vspace{-.2in}
\end{figure}

\subsection{Accessing and caching data}
\label{sec:access-data}

\myparagraph{Reads}
Figure~\ref{fig:data-access} shows the four ways to read data.
A {\em local hit} finds the data cached in the local \faast cachelet.
A {\em local miss} occurs when the local \faast cachelet is the owner for the object and does not currently cache the data.
The cachelet directly fetches the data from remote storage.
A {\em remote hit} occurs when the data misses the local cachelet but is found in the owner's cache.
Finally, a {\em remote miss} occurs when the access misses both the local cache and the owner's cache.
The owner fetches the data from remote storage and caches it locally.
In all cases, \faast cachelets cache objects locally, even if they are not the owners, for performance and locality.
Thus, a popular object will incur at most one remote hit per cachelet, and local hits thereafter (besides the optional consistency version check, described below).

\faast uses consistent hashing to determine object ownership.
We choose consistent hashing because (a) it avoids having to track object metadata (\eg list of objects in each instance), and (b) it reduces rebalancing as instances are added/removed: on average, only $numObjects / numInstances$ need to be remapped~\cite{aumala2019pasch,karger1997consistent}.
To maintain transparency, the object namespace is the same as that used by the remote storage service.
This design choice also enables efficient communication between cachelets, which as observed by prior work~\cite{fouladi2019gg,carreira2019cirrus,jonas2019cloud} is beneficial for applications such as ML training.

\myparagraph{Writes}
When the application needs to output data, \faast writes through to the owner cache.
The instance executing the function sends the data to the owner cache, which then writes it to remote storage.
This guarantees that the owner always has the latest version that the application has written.
By default, the write happens synchronously to the owner and synchronously to remote storage.
This offers high fault tolerance while trading off performance, since applications must wait until the write completes before proceeding with their execution.
Applications can {\em optionally} configure \faast to write asynchronously or not write to remote storage at all.
Because \faast is tied to each application, different applications can use different policies at the same time.

\myparagraph{Consistency}
Table~\ref{tab:consistency-versions} summarizes the possible read/write settings for \faast, and the performance, consistency, and fault tolerance (FT) they achieve.
By default, when reading an object, \faast first verifies the version in the cache matches the one in remote storage.  No data gets transferred during this check.
If the version matches (the common case), no object is retrieved.
This verification, combined with synchronous writes to remote storage, offers strong consistency (first row of Table~\ref{tab:consistency-versions}).
We set this as the default because it provides the same fault tolerance with better performance than the production FaaS platform.

Some applications may be willing to trade off consistency for performance (\eg, ML inference).
For those applications, \faast can read any cached version and write asynchronously to remote storage.
This weakens its fault tolerance, and provides only eventual consistency.
Applications can also completely skip writing to remote storage and rely on the distributed cache.
In Section~\ref{sec:consistencyeval}, we quantify the performance and consistency impact of these settings.

\begin{table}[t]
    \resizebox{\linewidth}{!}{%
    \centering
    \begin{tabular}{lll|lll}
      Write   & Write & Read   &             &             &   \\
      target  & mode  & target & Performance & Consistency & FT\\
        \hline
        Storage & Sync  & Storage & Low         & Strong      & High\\
        Owner   & Sync  & Owner   & Medium      & Strong      & Medium\\
        Owner   & Async & Owner   & Medium      & Eventual    & Medium\\
        Owner   & Sync  & Local   & High        & Weak        & Medium\\
        Local   & Sync  & Local   & High        & None        & Low\\
        Local   & Async & Local   & High        & None        & Low\\
    \end{tabular}
    } %
    \caption{
      Performance, consistency, and fault tolerance (FT) for different write/read settings.
      By default, \faast writes to storage synchronously, and reads the version from storage (first row).
    }
    \label{tab:consistency-versions}
    \vspace{-.2in}
\end{table}

\subsection{Pre-warming application data into \faast}
\label{sec:loadunload}

To \emph{pre-warm} future cachelets, \faast records metadata about the cache, off the critical path, when the Frontend unloads the application.
This includes the size of each cached object, its version, its number of accesses of each type (\eg, local hit, remote miss, produced as an output), and its average inter-arrival access time.
We timestamp each metadata collected with the unload timestamp to capture the state history of the cachelet.
As we describe next, this is necessary for applications that are rarely-invoked (\eg, once per hour), since their data access pattern cannot be determined by a single invocation.

\faast needs to decide when to load an application into memory.
For this, the Frontend leverages a previously\-/proposed hybrid histogram policy~\cite{shahrad2020serverless}.
The policy tracks the idle times between invocations of an application in a histogram.  When the application is unloaded, the Frontend uses the histogram to predict when the next invocation is likely to arrive, and schedules the reload of the application for just (\eg, 1 minute) before that time.  Our approach would work with any other cold-start prevention policy as well. 

At this point, \faast needs to decide what data objects should be loaded into the new cachelet.
To do so, it collects and merges the metadata across all cachelets over a pre-set period of time.
The period of time is based on the application's invocation frequency, which can be determined using the hybrid histogram policy.
Next, \faast determines the objects to be loaded using the following two conditions.
First, if an object's local or remote cache hit rate
is greater than a threshold,
the object should be loaded.
This indicates that the object has temporal locality.
Second, if an object is accessed more than once across the merged metadata, the object should be loaded.
This benefits rarely-invoked applications by loading objects accessed \emph{across} unload/load periods (\eg, an ML inference application's model and labels).
Once the objects to be loaded are determined, the \faast cachelet loads the objects that it owns based on consistent hashing (Section~\ref{sec:access-data}).

To avoid competing with on-demand accesses, \faast pre-warms the cache
only when the application is not executing, \ie before an invocation
arrives or right after a function execution ends.  If we cannot avoid
a cold-start invocation, the only data that is loaded into the cache is its inputs.

\subsection{Evicting application data from \faast} \label{sec:eviction}
The memory capacity of each application (and thus \faast) cachelet is set by the provider (typically a few GBs).
Cachelets do not consume any memory beyond that allocated to the application.

Each Memory Daemon monitors the memory usage of the function and the cachelet.
When the memory consumed by the function (\ie, heap memory) and the cachelet (\ie, cached objects) is within a small percentage (\ie, 10\%) of the application's total memory capacity, it evicts objects.

Eviction policies are often designed to cover the broad set of applications that can run on the platform~\cite{beckmann2018lhd,beckmann2017eva,berger2017adaptsize}. 
In contrast, as \faast is tied to an application, it can use \emph{per-application eviction policies}.
Hence, the eviction policy can be kept simple and tailored to an application's data access pattern as needed.

We implement two policies that we expect will work well for
many applications.  The first is a simple Least-Recently-Used (LRU)
policy that prioritizes the eviction of objects that are not owned by
the evicting cachelet.  Only after these objects are evicted, does the
policy consider owned objects in LRU order.
This is the default policy.
The second policy targets objects that are larger than a threshold (\eg, 12KB)
and not owned by the evicting cachelet.  If there are not enough of
these objects, the policy evicts non-owned objects smaller than
the threshold.  If we still need more capacity, we evict owned objects
that are larger than the threshold, before resorting to LRU for the
remaining ones.  In both eviction policies, targeting non-owned
objects first increases the number of remote hits, but also
minimizes the number of remote misses which are most expensive.
For the applications we consider, we find that targeting non-owned objects first improves
application performance by $\sim$20\% on average when
multiple cachelets are running.

Each of these eviction policies fits our emerging applications nicely:
ML inference matches the first policy and JupyterLess the second.  ML
inference applications that exhibit high invocation rates (\eg,
frequently used recommendation models~\cite{hazelwood2018facebook})
can quickly fill up a \faast cachelet's capacity with invocation
inputs (\eg, images) and outputs (\eg, labeled objects).  Across
invocations, only the model and labels are typically reused; inputs
and outputs change each time.  Thus, for ML inference and similar
applications, the first policy (LRU) is sufficient, since the inputs
and outputs will be evicted when the cachelet's capacity reaches its limit.

JupyterLess data objects can be classified into two types: (a) small
objects that maintain the notebook's state (\eg, a dictionary object)
and (b) larger objects that are used for data analysis (\eg, a
DataFrame).  A notebook's state is typically reused across
invocations, and should thus be cached as much as possible.  Larger
objects are reused less frequently and can be replaced more
aggressively.  Thus, the second policy (size-based) is appropriate.

\faast allows for future eviction policies beyond the ones described above.
For example, objects can be given a time-to-live (TTL) and get evicted when the TTL expires.

\subsection{Charging for \faast} \label{sec:charging}
When using \faast, we expect FaaS providers to charge users only for the memory of the accessed data and not all the cached objects.
FaaS providers should also not charge for pre-warming metadata in the same manner that they do not charge for function metadata (\eg function registration).

\vspace{-.1in}
\section{Scaling \faast}
\label{sec:scaling}

FaaS platforms typically include a Scale Controller responsible for
scaling applications in/out (Figure~\ref{fig:arch-diagram}).  As it is
part of the front-end component, the Scale Controller monitors the
end-to-end performance and the load offered to each application.  It
also periodically queries the FaaS runtime running each application
instance for a vote on how many more instances to add: a positive
number means a vote to scale-out and a negative number means a vote to
scale-in.  Based on the information for an application, it makes a
scaling decision and effects it.  \faast extends this mechanism by
including cache-specific metrics when deciding on how to vote.  We
also extend the Scale Controller to accept unrequested votes, when
scaling is needed immediately.  When the controller adds or removes
an application instance, \faast reassigns the objects' ownership using the
Membership Daemon's consistent hashing.

\faast has three types of scaling:

\myparagraph{Compute scaling} FaaS platforms scale the number of
application instances based on its rate of incoming requests, its
number of in-flight requests (queue sizes), and/or its average
response time.  Degrading performance, high request rates, or long
queues cause scale-out; the opposite causes scale-in.  Since every
application instance includes both compute and caching resources, this
traditional way of scaling is sufficient.

\myparagraph{Cache size scaling} \faast also scales to match the
application's working set size.  For example, a JupyterLess notebook
performing data-intensive operations may not fit the
entire working set in the cache, leading to a high eviction rate.  To
address this, each cachelet tracks the number of evictions of
each locally-cached object and votes to scale out by one instance, if
any object has been evicted more than once since the last controller
query.  If no object has been evicted more than once but there is
still substantial cache access traffic, \faast votes to do nothing
(add 0 instances).  It votes to scale in by one instance when the
frequency of accesses is low.

Many existing caching systems statically allocate resources and either
cannot auto-scale their capacity as the amount of data accessed varies
or require application hints to do so.
OFC uses per-application machine learning models to achieve the same dynamic cache size scaling, which requires frequent retraining and mechanisms to prevent application ``out-of-memory'' failure.

\myparagraph{Bandwidth scaling} \faast also supports applications with
large input objects.  For such objects, \faast equally partitions the
download from remote storage across multiple cachelets to (a) create a
higher \emph{cumulative} I/O bandwidth to remote storage, and (b)
exploit the higher communication bandwidth \emph{between} 
instances ($BW_{Inst}$) compared to the bandwidth between each
instance and remote storage ($BW_{BS}$).

When a \faast cachelet receives an object access, it iteratively computes the data transfer latency, $T_{DR}$, for a number of instances $N$ (starting at the current number) and the object size $S$ using the following formula:
$$T_{DR} = T_{Load} + S/N \times 1/BW_{BS} + (S - S/N) \times 1/BW_{Inst}$$
where $T_{Load}$ is the instance loading latency.
\faast periodically profiles $T_{Load}$, $BW_{BS}$, and $BW_{Inst}$ to account for variations in the network and the remote storage bandwidths.
The iterative process stops at the $N$ where $T_{DR}$ changes by less than 10\% or when $T_{DR}$ increases between iterations.
If the resulting $N$ is greater than the current number of instances, the \faast
cachelet {\em immediately} contacts the controller to scale out to $N$.
\faast then waits for the new instances to be created (by checking the
Membership Daemon) and sends each of them a request to download a
different byte range of size $S/N$.  As scale-in is not as
time-critical, \faast does it through periodic voting (when queried by
the controller) as the number of object accesses becomes small.

We find that bandwidth-based scale-out is worthwhile for objects on
the order of hundreds of MB (Section~\ref{sec:scaleeval}); this will
become smaller as cold-start optimizations continue to
appear~\cite{agache2020firecracker,oakes2018sock,du2020catalyzer, cadden2020seuss}.
Existing systems do not support bandwidth scaling, and instead rely on
the user to determine the right number of chunks and instances.

\myparagraph{Handling conflicting scaling requests} The scaling
policies work concurrently, so there may be scenarios where they make
conflicting scaling requests to the Scale Controller.  For example,
compute scaling may want to scale out, while cache size scaling may
want to scale in.  When there are conflicting votes, the controller
scales out if \emph{any} policy determines that scale-out is needed.
It scales in if \emph{all} policies suggest scale-in will not hurt.
This is similar to the approach taken by existing systems for
right-sizing storage clusters~\cite{klimovic2018pocket}.

\myparagraph{Idle function computation resources}
When instances are added due to cache size or bandwidth scaling, their computation resources can be wasted. 
\faast minimizes resource waste by scaling in when the frequency of accesses is low.
Providers can also leverage resource harvesting~\cite{zhang2016harvesting} to run low-priority tasks (\eg, analytics jobs) on these resources when they are not in use.

\section{Implementation} \label{sec:implementation}

We implement \faast for a large-scale FaaS platform used in production,
and have open-sourced the bulk of it~\cite{cache-github-binding,cache-github-shmem}.

\myparagraph{Production FaaS platform}
In our platform, a user application comprises one or more functions.
Each function defines its data bindings, which \faast uses to transparently load and manage objects: trigger (\eg, HTTP request), inputs (\eg, blob), and outputs (\eg, message queue).
Users optionally set \faast policies (scaling, eviction, and consistency) using simple application-specific configurations at deployment time.

\begin{figure}[t]
  \centering
  \includegraphics[width=\columnwidth]{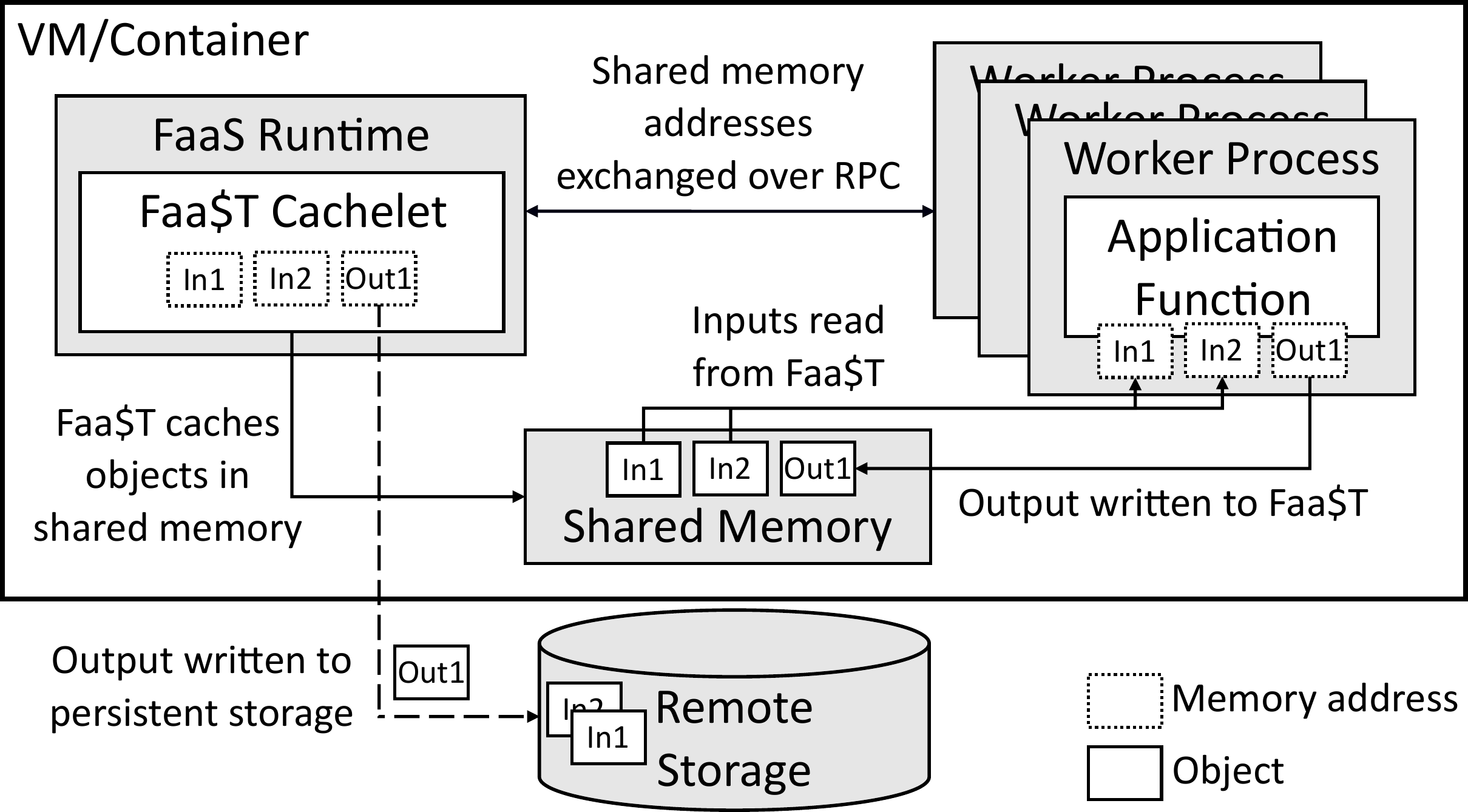}
  \caption{
    \faast integrated with the FaaS platform.
    Runtime and workers point to the same shared memory objects.
  }
  \label{fig:runtimeandworker-diagram}
  \vspace{-.2in}
\end{figure}

As we show in Figure~\ref{fig:runtimeandworker-diagram}, an application instance executes in either a VM or a Docker container, and includes the FaaS runtime and function-execution workers.    
Upon receiving incoming requests (\eg, as a result of an incoming HTTP request, a new blob being created), the runtime collects the requested input bindings and invokes the function in a worker while passing the appropriate arguments to it.
When the worker finishes executing the function, it replies to the runtime with the produced output(s) and the runtime processes them (\eg, writes a blob to remote storage or writes to a message queue).
If there are multiple concurrent invocations, more worker processes can be spawned on the same instance to execute them in parallel.
The platform leverages an existing remote storage service that is not tailored to FaaS.

As Figure~\ref{fig:arch-diagram} shows, a Frontend component handles HTTP requests and does compute scaling.  We extend this component to implement bandwidth-based scaling (Section~\ref{sec:scaling}).

\myparagraph{Caching data}
We implement the core of \faast in the runtime (C\# code) with minimal changes to the workers (Python and Node.js).
In the original design, the runtime and workers exchanged control and data messages over a persistent RPC channel.
\faast replaces the data messages with a shared memory area, while keeping control messages over RPC.  The shared area is also where \faast caches data.
Data communication between the \faast cachelets and the workers happens via passing shared memory addresses, reducing end-to-end latency.
In addition, unlike existing systems that need to maintain duplicate object copies~\cite{mvondo2021ofc}, using shared memory reduces the memory footprint by only keeping a single object copy~\cite{cache-github-shmem}.
The workers managed by the same runtime share the cached objects.

When the runtime prepares input data bindings before invoking a function,
\faast intercepts them and checks the cache first
(Section~\ref{sec:access-data}).  When a function produces
an output, \faast caches it for future invocations.  This cache write
triggers any functions that have the newly added object as their
trigger binding.
This improves latency for applications
that rely on writing intermediate outputs to external sources (\eg,
blob storage) to trigger subsequent functions.

We support applications written in C\#, Python, and Node.js.
Supporting other languages would require minimal changes.
We use the shared memory APIs already available in most languages for both Linux and Windows.
When we run applications in containers (vs VMs), we share (setting up permissions) the cache space across containers.

\myparagraph{Distributed cache}
Each runtime instance saves some metadata
about the running application in a blob from remote storage.
We store the \faast membership information in this blob and the \faast cachelets periodically heartbeat their state there.
The consistent hashing algorithm uses SHA256 for hashing and 100 cachelet replicas for load balancing.
More replicas did not improve load balancing and increased the ownership lookup time.
Fewer replicas created ownership hot spots.

As our platform already uses HTTP for communication between its components, we use this interface to exchange data between \faast cachelets.
We evaluated other approaches like RPC (with Protocol Buffers~\cite{protobuf}) but the improvements were negligible and the complexity of maintaining a new channel would offset them.
We could also leverage RDMA-based communication but have not experimented with it.

\myparagraph{Other platforms}
The design and implementation of \faast is extensible to other FaaS platforms.
Most platforms have a similar architecture and \faast directly applies to the equivalent components (\eg, runtime, worker processes, Scale Controller).
All platforms use the concept of triggers and interact with external data services.
However, not all of them use bindings to map the data but rely on libraries to explicitly access it inside the function body.
We would need to extend these libraries (\eg, Boto3~\cite{aws-boto3} in AWS Lambda) to interact with \faast and look-up the cache before accessing the remote storage.
These extensions would be equivalent to modifying the binding process in our platform.

\section{Evaluation}
\label{sec:evaluation}

\begin{figure*}[t]
  \centering
  \begin{subfigure}[t]{0.33\linewidth}
    \includegraphics[width=1\textwidth]{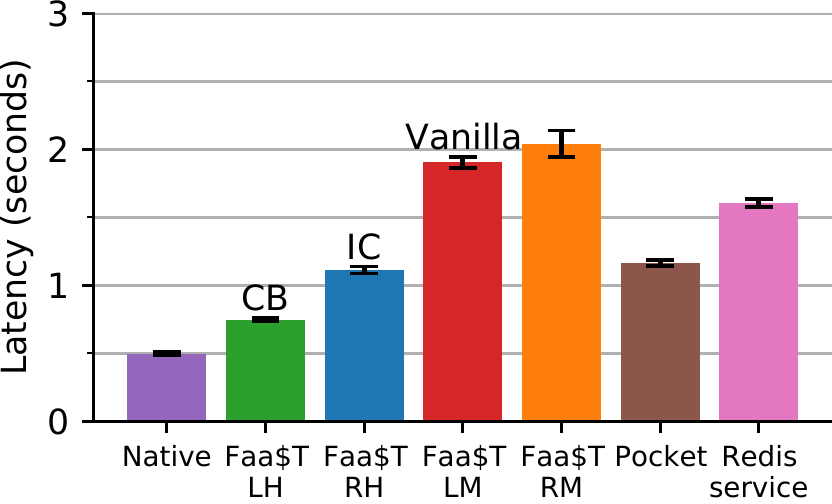}
    \caption{AMBER Alert pipeline}
    \label{fig:inf-pipeline}
  \end{subfigure}
  \begin{subfigure}[t]{0.33\linewidth}
    \includegraphics[width=1\textwidth]{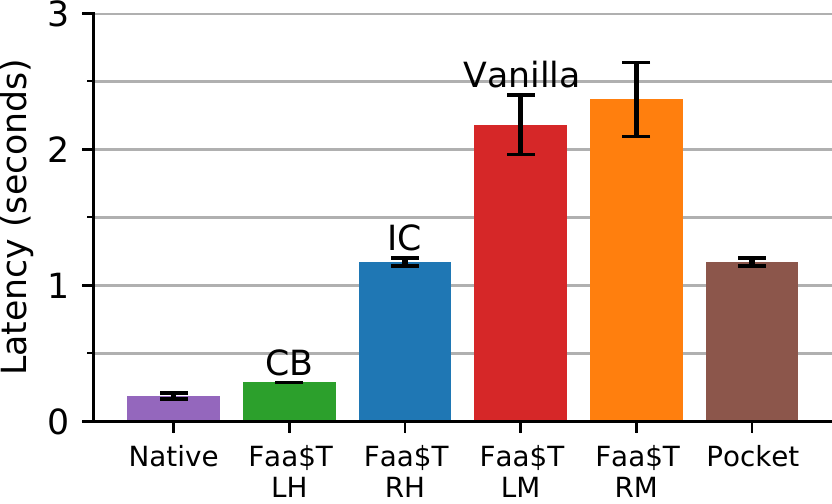}
    \caption{AlexNet (239MB)}
    \label{fig:inf-alexnet}
  \end{subfigure}
  \begin{subfigure}[t]{0.33\linewidth}
    \includegraphics[width=1\textwidth]{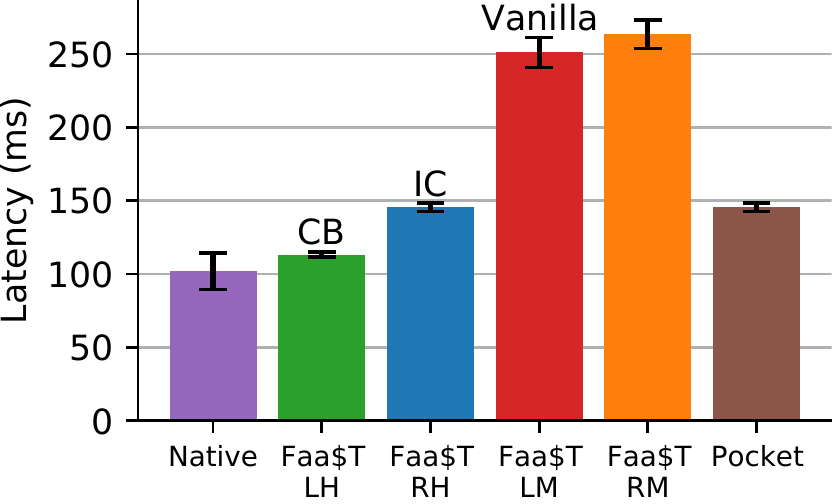}
    \caption{SqueezeNet (5MB)}
    \label{fig:inf-squeezenet}
  \end{subfigure}
  \caption{
    \faast versus existing systems for ML inference.
    \faast improves application performance by accessing data in local or remote cache instances.
    (LH = Local Hit, RH = Remote Hit, LM = Local Miss, RM = Remote Miss, CB = Cloudburst, IC = InfiniCache.)
  }
  \label{fig:inf-all}
  \vspace*{-.15in}
\end{figure*}

\subsection{Methodology}

\myparagraph{Comparison points}
We perform two types of comparisons.
The first is an analysis of running the application traces from Section~\ref{sec:character} on top of \faast.
This allows us to show the improvements these applications would get with \faast.

The second evaluates the four access scenarios that functions may encounter: objects are accessed through local hits (LH), local misses (LM), remote hits (RH), or remote misses (RM).

We compare \faast against six baselines for performance and cost:
(a) a large, local VM where all accesses are local and there are no function invocation overheads ({\it Native});
(b) our commercial FaaS offering ({\it Vanilla}) without \faast that accesses all objects from remote object storage. Its performance is equal to that of \faast LM;
(c) InfiniCache ({\it IC})~\cite{wang2020infinicache} that we approximate by statically configuring \faast to use only remote instances. Its best case performance is equal to that of \faast RH; 
(d) Cloudburst's caching layer ({\it CB})~\cite{sreekanti2020cloudburst}. Its best case performance is equal to that of \faast LH;
(e) {\it Pocket}~\cite{klimovic2018pocket}, approximated with a manually managed Redis VM with all data available at memory speed (no Flash accesses); and
(f) a commercial Redis service ({\it Redis service}).
The Redis service is the offering that matches our VM size in memory and network bandwidth.
It is akin to what is used by Locus~\cite{pu2019locus}.
Data is stored and accessed from Redis as opposed to remote object storage.

\myparagraph{Applications}
We use the two applications from Section~\ref{sec:newapps}: ML inference and JupyterLess notebooks.
We use application latency and cost as primary metrics.
For each experiment, we report the mean and standard deviation of 3 runs.

For ML inference, we evaluate both single model inference and inference pipelines.
For single model inference, we use two separate models that differ in latency and resource footprints: SqueezeNet~\cite{iandola2016squeezenet} (5MB), and AlexNet~\cite{krizhevsky2012imagenet} (239MB).
For the inference pipeline, we evaluate the AMBER Alert pipeline of Figure~\ref{fig:mlpipeline-diagram}; the output of the bounding box model (MobileNet Single-Shot Detector~\cite{howard2017mobilenet}, 35MB) is fed into people (ResNet50~\cite{he2015resnet50}, 97MB) and car recognition (SqueezeNet, 5MB) models.
In all inference cases, functions access an input image, the model, and the class labels (a text file).

For JupyterLess, we use five notebooks: (a) single message logging (No-Op); (b) summing a 350MB DataFrame column; (c) capacity planning with data collection and plotting; (d) FaaS characterization of Section~\ref{sec:character}; and (e) counting up to 1K.
The function data objects consist of the notebook state after each cell's execution, which is stored in JSON format.

\myparagraph{Experimental setup}
Each application instance is a single VM; the default instance size in our experiments includes 8vCPUs with 28GiB of DRAM and up to 500MB/s network bandwidth.  I/O bandwidth to remote storage is lower at 90MB/s for large objects.
They run Ubuntu 18.04 with 5.4.0 kernel on Intel Xeon E5-2673 CPUs operating at 2.40GHz.
In our production setting, the VMs are pre-provisioned: an application instance cold-start involves loading and deploying the serverless runtime together with the application code.

\myparagraph{Cost model}
We derive user costs following the common pricing by FaaS and cloud providers. Function invocations are charged for the time and the resources they take ($\$/GB\mathit{\mhyphen}s$, order of $10^{-5}$), while VMs are charged for their lifetime ($\$/s$, order of $10^{-1}$).
We assume \emph{Native} and systems with additional resources are statically provisioned the whole time.  
Specifically, we charge for extra resources whenever the caching or storage system is external to the FaaS platform (\ie, Pocket, Redis, Redis Service) or specialized for FaaS in some way (CloudBurst).  
Except for Redis service, we charge these systems for one extra VM of the same instance size as the default application instance.  The VMs are charged their on-demand prices.
For Redis service, we provision per service class and charge the class's price.
The additional resource costs can be amortized by multiple applications sharing the same resources.  Vanilla, InfiniCache, and \faast use existing commodity storage (\eg, AWS S3, Azure Storage), so we do not charge them for extra resources.
We also add the cost of remote storage data transfer ($\$/op$, order of $10^{-6}$) to \faast LM and RM, and Vanilla.

\begin{figure}[t]
  \centering
  \includegraphics[width=\columnwidth]{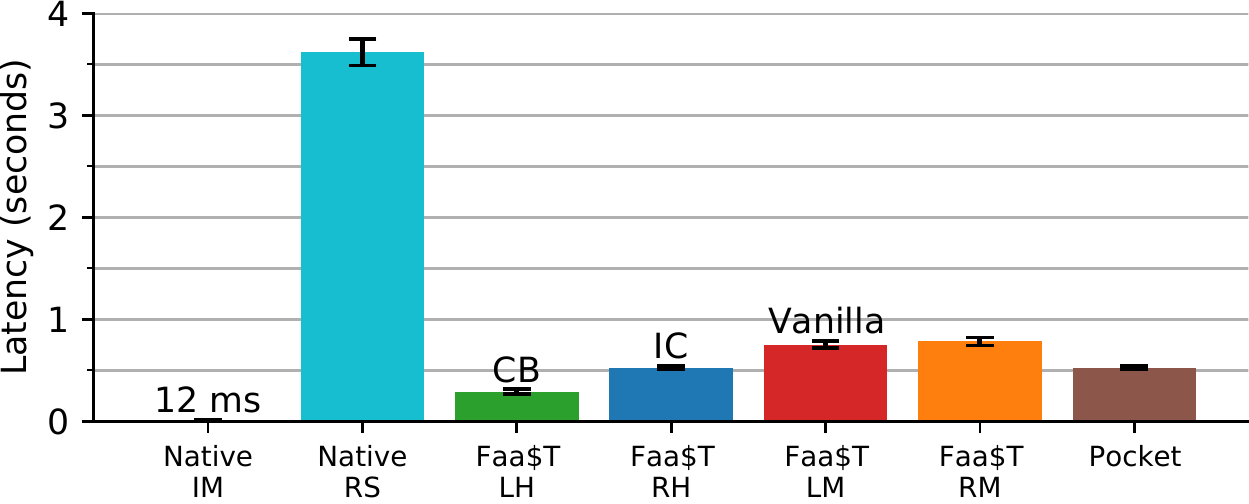}
  \caption{
    Performance of summing a 350MB DataFrame column in a JupyterLess notebook.
    There are two Native setups: In IM the DataFrame is already pre-loaded before summing, and RS fetches it from remote storage before summing.
    \faast improves application performance by accessing data in local or remote cache instances.
  }
  \label{fig:parquet}
  \vspace{-.15in}
\end{figure}

\subsection{\faast with applications run in production} \label{sec:prodappeval}

We simulate the end-to-end performance of the \faast applications from Section~\ref{sec:character}.
Our simulator uses the default policies for consistency (synchronous writes to owner, synchronous writes to remote storage, read version from remote storage) and eviction (LRU), and implements the scaling policies described in Section~\ref{sec:scaling}.
To model \faast's access latencies, we measured read and writes latencies for 1B to 2GB object sizes using our \faast implemention described in Section~\ref{sec:implementation}.
We vary the size of \faast from 1KB to 128MB; larger cache sizes showed no further improvement.
We also vary the unload period and show how it affects performance when \faast cannot pre-warm frequently-accessed objects.

\begin{figure}
  \centering
  \begin{subfigure}[t]{0.49\linewidth}
    \includegraphics[width=1\columnwidth]{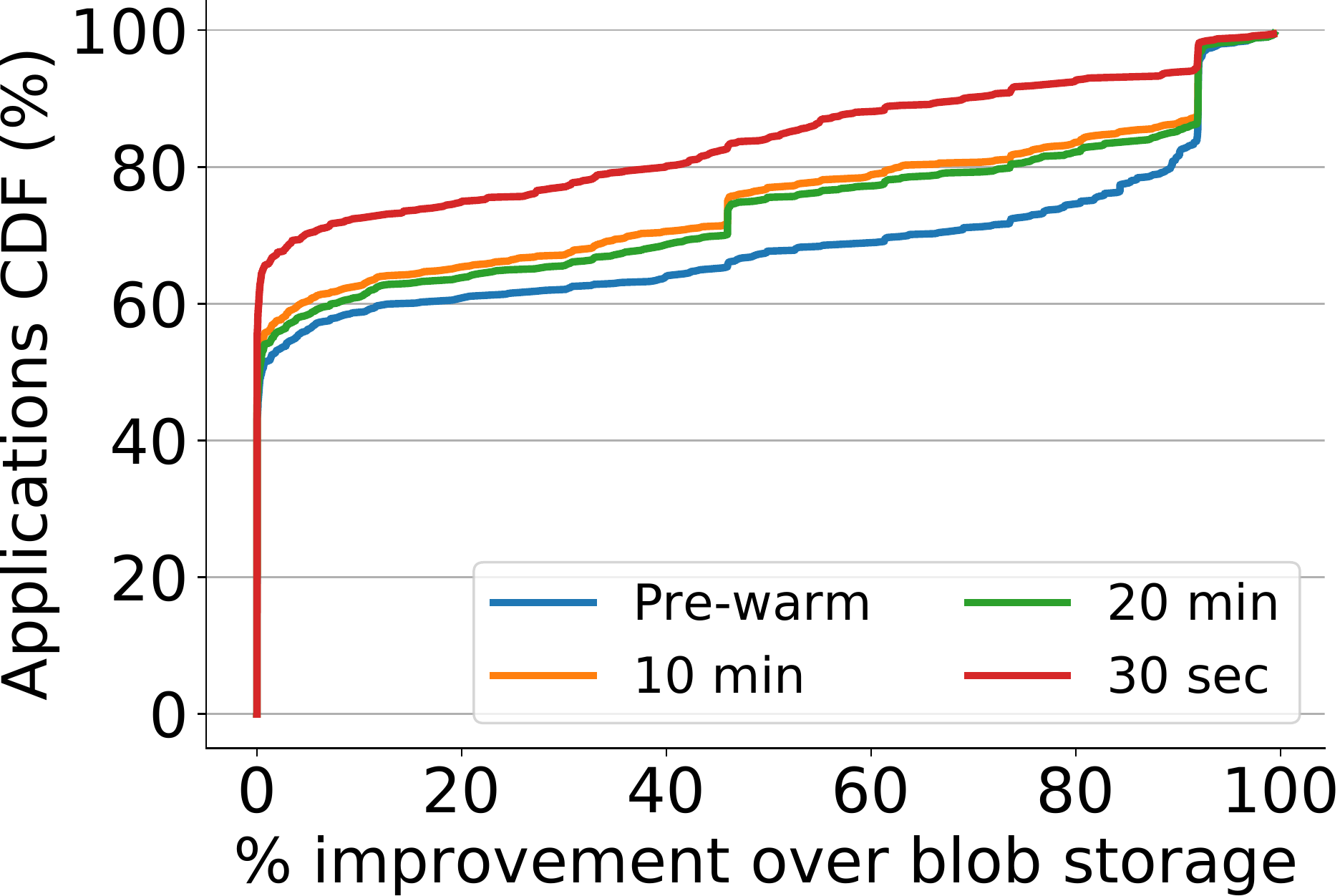}
  \end{subfigure}
  \begin{subfigure}[t]{0.49\linewidth}
    \includegraphics[width=1\columnwidth]{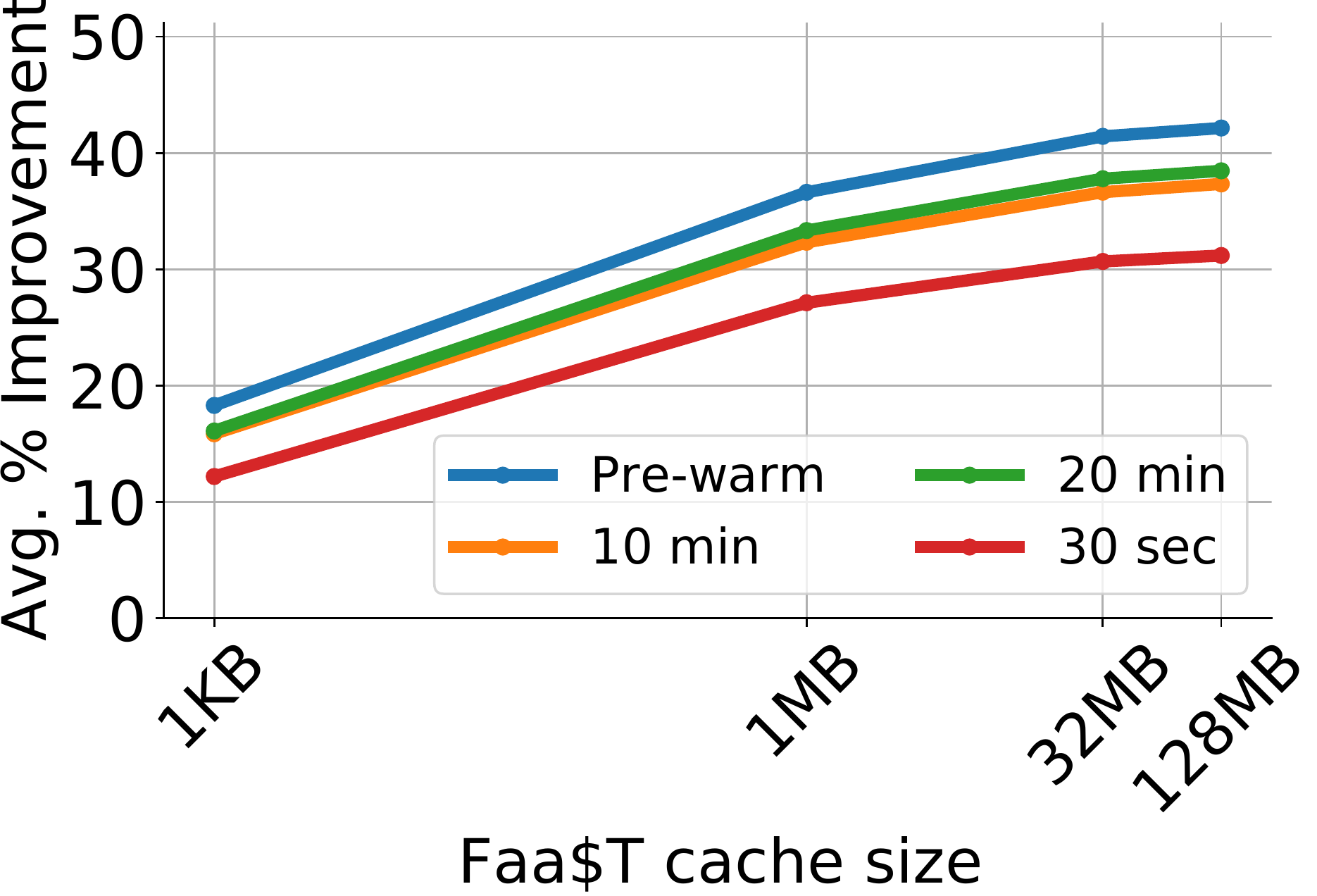}
  \end{subfigure}
  \caption{
    CDF of percent improvement over blob storage for a 128MB cache (left) and average percent improvement as the size of \faast varies (right) with the application traces from Section~\ref{sec:character}.
    30 sec, 10 min, and 20 min represent different unload periods with pre-warm disabled.
    \faast with pre-warm improves performance by 50\% or more for about 35\% of applications, and has an average improvement of over 40\% for a 128MB cache.
  }
  \label{fig:sim-apps}
  \vspace{-.05in}
\end{figure}

Figure~\ref{fig:sim-apps} shows the CDF of percent improvement over blob storage for a 128MB cache (left) and average percent improvement as the size of \faast varies (right).
First, with just 128MB, \faast with pre-warm improves performance by 50\% or more for about 35\% of applications.
\faast also has an averge improvement of over 40\% for a 128MB cache.
Second, as the unload period gets smaller, \faast's pre-warm becomes more important to ensure frequently-accessed objects are available during the next application invocation.
Third, improvement is correlated with reuse: we found that smaller objects tend to be reused more often, which resulted in greater performance improvements.
Finally, we note that \faast is designed to support applications that run in production today (with object sizes of tens to hundreds of KB), but also for future applications that will access much larger objects (hundreds to thousands of MB) as shown in Section~\ref{sec:existingeval}.

\subsection{Comparing \faast to existing systems} \label{sec:existingeval}
\myparagraph{ML inference}
Figure~\ref{fig:inf-all} shows the latency for the AMBER Alert pipeline and single inference with AlexNet and SqueezeNet.
First, the figure shows that \faast LH improves latency by 50\%, 87\%, and 60\% faster than Vanilla for the AMBER Alert pipeline, AlexNet, and SqueezeNet, respectively.
This demonstrates that avoiding remote storage accesses and using cache triggers can significantly improve FaaS performance.
Second, for the AMBER Alert pipeline, \faast's LH and RH are faster than using a Redis service, while \faast RH is equivalent to using a manually managed Redis VM (Pocket in Figure~\ref{fig:inf-pipeline}).
This is significant given the complexity of manually managing a Redis VM and the significant cost of using a Redis service (discussed below).
\faast offers lower latency, while remaining transparent and relieving users of any configuration burden.
Third, \faast's LM and RM exhibit similar latencies, with the variability coming from the accesses to remote storage.
This suggests using a multi-instance \faast cache does not further penalize cache miss performance.
Finally, \faast LH and RH perform well for both small (input images and class labels) and large objects (the models).

\myparagraph{JupyterLess}
Figure~\ref{fig:parquet} shows the performance of summing a 350MB DataFrame column in a JupyterLess notebook.
There are two Native setups: for In-Memory (IM) the DataFrame is pre-loaded in memory before summing, while for Remote Storage (RS) the latency of remote storage access is counted as part of the summation.
Similar to the ML inference applications, \faast LH and RH improve performance compared to Vanilla by 62\% and 29\%, respectively.
Compared to Native RS, \faast's LH and RH improve performance by 92\% and 86\%, respectively.

Table~\ref{tab:jupyterlessapps} shows the latency of running a capacity planning notebook (includes data collection and plotting), the FaaS characterization of Section~\ref{sec:character}, and a No-Op notebook that logs a single message.
Each run has a mix of data access scenarios for \faast, since a local copy of the Jupyter state is saved per cell, and is read from a cachelet when executing the following cell.
For the former two notebooks, the performance gap with Native IM comes from serializing plots and sending them to the notebook user interface.
These results show that \faast can run JupyterLess notebooks interactively, and with near-native performance when the notebook is not trivial.

\begin{figure}
  \centering
  \begin{subfigure}[t]{0.49\linewidth}
    \includegraphics[width=1\columnwidth]{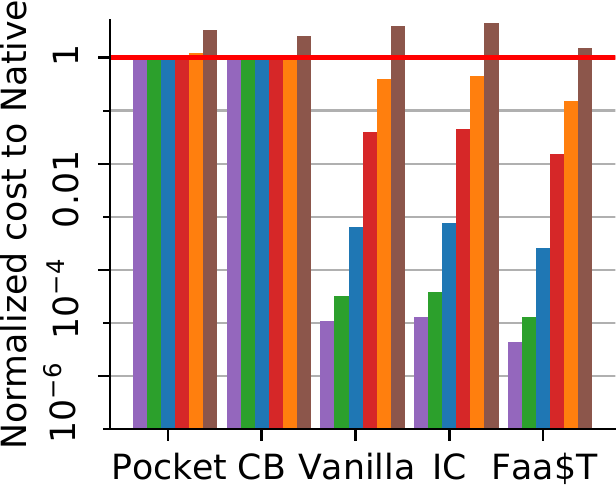}
    \caption{AMBER Alert pipeline}
    \label{fig:inf-cost}
  \end{subfigure}
  \begin{subfigure}[t]{0.49\linewidth}
    \includegraphics[width=1\columnwidth]{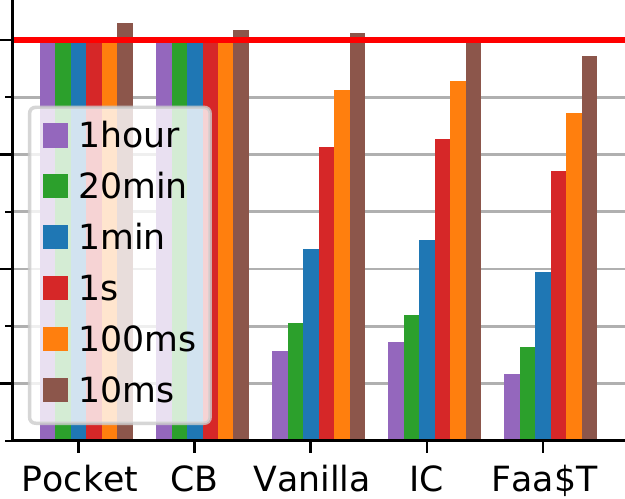}
    \caption{350MB DataFrame notebook}
    \label{fig:parquet-cost}
  \end{subfigure}
  \caption{
    Cost normalized to Native VM for ML pipeline and JupyterLess notebook.
    Each bar represents an inter-invocation period.
    The y-axis is in log-scale (lower is better).
    \faast is 50\% to 99.999\% cheaper than the baselines.
  }
  \label{fig:cost-all}
  \vspace{-.05in}
\end{figure}

\begin{table}[t]
    \resizebox{\linewidth}{!}{%
    \centering
    \footnotesize
    \begin{tabular}{l|rr}
        Notebook & Native IM & \faast \\ \midrule[1pt]
        Capacity planning & 6.0s $\pm$ 0.2s & 8.0s $\pm$ 0.3s \\
        FaaS characterization & 40.4s $\pm$ 8.9s & 68.0s $\pm$ 0.8s \\
        No-Op & 0.2ms $\pm$ 0ms & 34.4ms $\pm$ 8.7ms \\ \bottomrule
    \end{tabular}
    } %
    \caption{
      End-to-end latency running notebooks.
      \faast can run JupyterLess notebooks interactively.
    }
    \label{tab:jupyterlessapps}
    \vspace{-.15in}
\end{table}

\myparagraph{Cost}
Figures~\ref{fig:inf-cost} and~\ref{fig:parquet-cost} show the cost-per-hour comparison of running the applications end-to-end (lower is better).
Cost is normalized to that of {\it Native}, and the y-axis is in log scale.
We show cost for different application invocation IaTs between 10ms and 1 hour.
For context, the application with the median IaT in our characterization (Section~\ref{sec:character}) has an average IaT of 20 minutes.
For \faast, we show the case where the application ran end-to-end with all local hits.
We do not show Redis service due to its large cost (\faast is 6 order of magnitude cheaper).

The figure shows that \faast can provide huge cost savings.
For applications with infrequent invocations (\eg, once per hour), \faast is 99.999\% cheaper than a Native VM.
Per hour, \faast is cheaper than all baselines for all invocation intervals, except for the AMBER Alert pipeline at 10ms, where it is 33\% more expensive than Native.
For all other cases, \faast is 50\% to 99.999\% cheaper than caching and storage layers that require separate servers, such as Cloudburst and Pocket.
From Section~\ref{sec:character}, 99.88\% of applications have average IaT $\geq$
10ms: in these cases, the cost of servers would
almost always completely dominate the overall cost. 

\myparagraph{Discussion on comparison to Native}
Even when all accesses are served with local (LH) or remote hits (RH), \faast is slower than a Native VM with all data stored locally and there are no function invocation overheads.
However, as we have shown in Figure~\ref{fig:cost-all}, such a Native setup can be orders of magnitude more expensive than \faast, since we must keep all VMs running even when applications are idle.
Moreover, the user is responsible for resource and data management.
With \faast being a part of the FaaS runtime, users only pay for the time resources are consumed for both compute and caching.
Moreover, local hits are on the order of hundreds of ms, which is close to interactive for many use cases.

\begin{figure}[t!]
  \centering
  \includegraphics[width=\columnwidth]{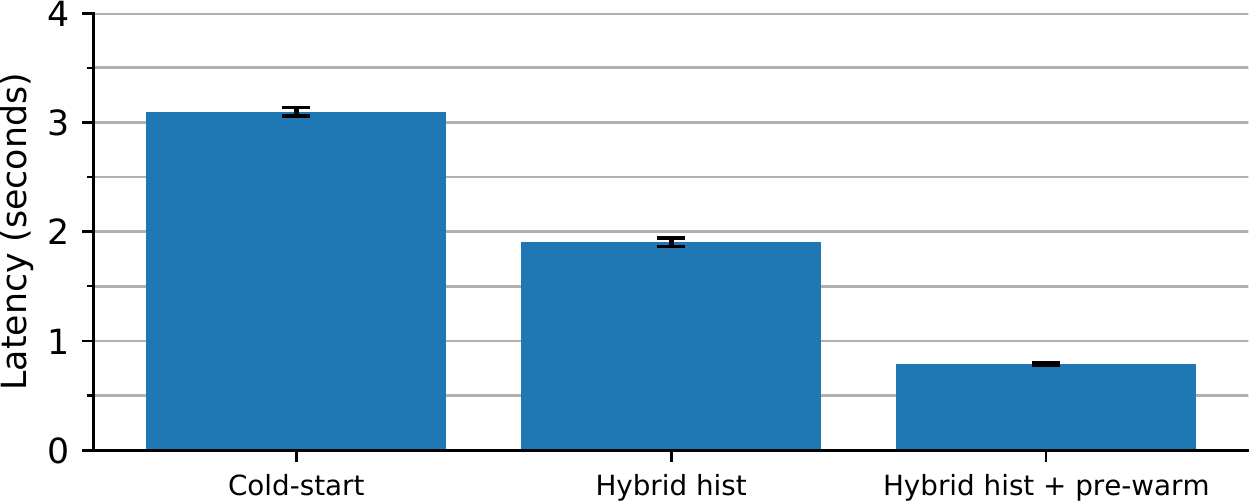}
  \caption{
    AMBER Alert pipeline performance:
    (a) instance loaded and \faast pre-warms based on past history (\texttt{Hybrid hist + pre-warm}),
    (b) instance loaded but not pre-warmed (\texttt{Hybrid hist}), and
    (c) instance not loaded (\texttt{Cold-start}).
    \faast automatically loads objects with spatial and temporal locality to improve latency.
  }
  \label{fig:dataloading}
  \vspace{-.15in}
\end{figure}

\begin{figure*}
  \centering
  \begin{subfigure}[t]{0.24\linewidth}
    \includegraphics[width=1\textwidth]{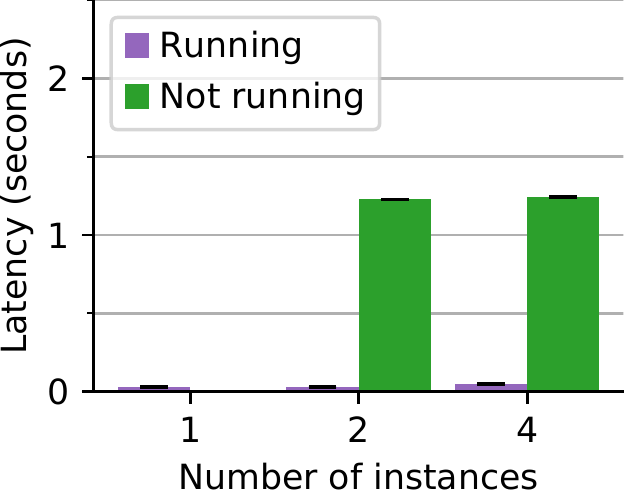}
    \caption{400KB}
    \label{fig:increasebw-400k}
  \end{subfigure}
  \begin{subfigure}[t]{0.24\linewidth}
    \includegraphics[width=1\textwidth]{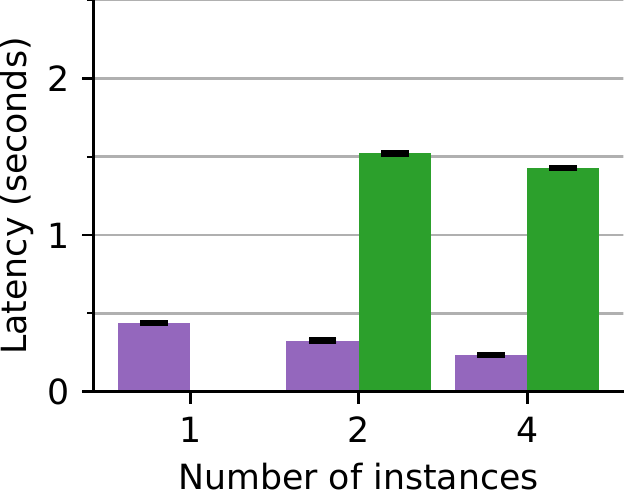}
    \caption{40MB}
    \label{fig:increasebw-40m}
  \end{subfigure}
  \begin{subfigure}[t]{0.24\linewidth}
    \includegraphics[width=1\textwidth]{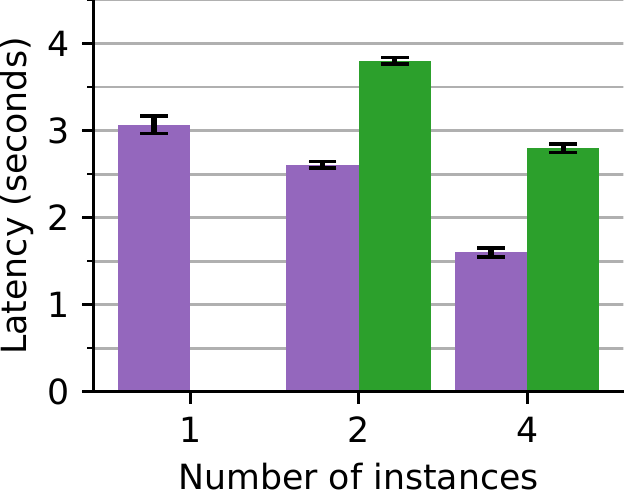}
    \caption{400MB}
    \label{fig:increasebw-400m}
  \end{subfigure}
  \begin{subfigure}[t]{0.24\linewidth}
    \includegraphics[width=1\textwidth]{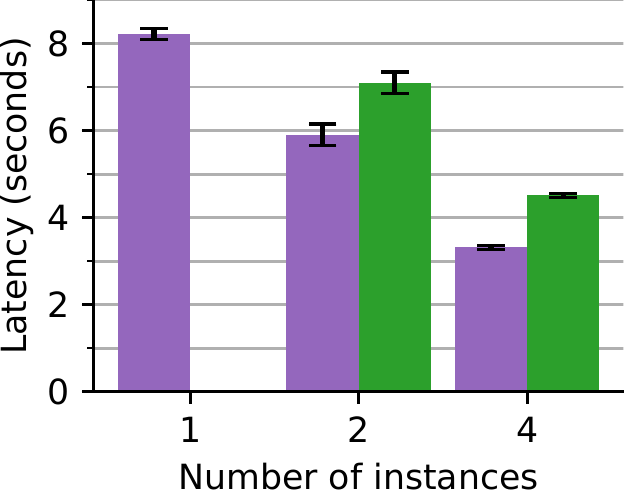}
    \caption{800MB}
    \label{fig:increasebw-800m}
  \end{subfigure}
  \caption{
    Latency of fetching an object from remote storage as the number of instances vary for increasingly large object sizes.
    If the instances are not loaded, they incur a cold-start; we only consider the running case for one instance.
    \faast determines whether to scale data loading across multiple instances to increase bandwidth.
  }
  \label{fig:increasebw-all}
  \vspace{-.15in}
\end{figure*}

\vspace{-.1in}
\subsection{Is \faast pre-warm effective?} \label{sec:prewarmeval}
In Section~\ref{sec:prodappeval}, we showed pre-warming is an important feature for improving existing application performance over remote storage.
We now use the AMBER Alert pipeline to evaluate the effectiveness of \faast data pre-warming under three scenarios: the application instance was loaded before the function invocation using the hybrid histogram policy~\cite{shahrad2020serverless} and \faast automatically pre-warmed the three models and three labels (135MB total) based on history (\texttt{Hybrid hist + pre-warm}); the instance was loaded before the invocation using the hybrid histogram policy, but not pre-warmed with any objects (\texttt{Hybrid hist}); and the instance was not loaded before the invocation (\ie, the runtime is not deployed; \texttt{Cold-start}).

Figure~\ref{fig:dataloading} shows the performance of the pipeline for these versions.
\faast's data pre-warming improves latency by 58\% and 74\% over no pre-warming and cold-start, respectively.
This is especially important if the AMBER Alert pipeline is infrequently invoked.

\begin{table}[t]
    \resizebox{\linewidth}{!}{%
    \centering
    \footnotesize
    \begin{tabular}{l|cc}
        Scenario & Heap growth succeeded? & Latency \\ \midrule[1pt]
        No Mem Daemon & No & 235.0ms $\pm$ 3.2ms \\
        Mem Daemon, no scaling & Yes & 678.6ms $\pm$ 64.8ms \\
        Mem Daemon, scaling & Yes & 502.9ms $\pm$ 28.5ms \\ \bottomrule
    \end{tabular}
    } %
    \caption{
      Latency of running a 350MB DataFrame summation in a JupyterLess notebook after growing heap memory, and whether the heap growth succeeded.
      We evaluate three scenarios: (a) without the Memory Daemon, (b) with the Memory Daemon, but no cache size scaling, and (c) with Memory Daemon, and \faast scales to two instances.
      \faast ensures application functionality is not compromised, and improves performance by scaling.
    }
    \label{tab:heapmem}
    \vspace{-.15in}
\end{table}

\subsection{Can \faast manage memory effectively?} \label{sec:managemem}
We consider the JupyterLess notebook application that sums a 350MB column.
After loading in the DataFrame and performing the summation, the application allocates an array that consumes 96\% of the application's total memory.
Then, the application again computes the summation of the 350MB column, which requires the reloading of the DataFrame that is evicted.
We show three scenarios:
(a) without the Memory Daemon to evict objects when the heap memory grows,
(b) the Memory Daemon evicts objects, but there is no cache size scaling,
and (c) the Memory Daemon evicts objects, and \faast scales to two instances.

Table~\ref{tab:heapmem} shows the performance of the second summation of the 350MB column, and whether the array allocation completed successfully.
When the Memory Daemon does not trigger object eviction, the array allocation fails, but the summation matches the performance of a local hit.
When the Memory Daemon triggers object eviction, but cannot scale, the array allocation succeeds, and the summation performance is a mix of local hits and misses.
Finally, when the Memory Daemon triggers object eviction, and scales the cache size to improve the number of remote hits, the array allocation succeeds, and the summation performance is a mix of all four data access types.
Since \faast opportunistically uses application memory, functionality is not compromised for applications that use large heap memory amounts.

\vspace{-.1in}
\subsection{Scaling as the object size varies} \label{sec:scaleeval}
\faast can scale the number of instances based on object sizes.
We consider four object sizes: 400KB, 40MB, 400MB, and 800MB. 
The amount of data downloaded by each \faast cachelet is evenly split between the available instances. For example, if there are two instances and the object size is 400KB, each one downloads 200KB.
The data is then processed at a single instance.
For each object size, we show two cases: (a) when all application instances are running, and (b) when additional instances (more than one) must be loaded in order to fetch the objects (thus incurring a cold-start latency).

Figure~\ref{fig:increasebw-all} shows the time to download the increasingly large object sizes from remote storage as the number of instances varies.
For small objects (400KB), the download latency is low enough that using more than one instance degrades performance, especially if the instances need to be loaded.
For the 40MB object, partitioning the download across multiple instances is beneficial if the instances are already loaded.
It is 47\% faster to download the object with four cachelets than with one.
For the 400MB object, it is 9\% faster to download the object with four instances than one if the instances are not already loaded, and 47\% faster to download the objects with four instances than one if the instances are already loaded.
For the 800MB object, it is beneficial to use four cachelets to load the object in parallel, even if the instances need to be loaded.
If the instances are already loaded, it is 60\% faster to download with four cachelets compared to one.
If the instances need to be loaded, it is 44\% faster.

\begin{table}[t]
    \resizebox{\linewidth}{!}{%
    \centering
    \begin{tabular}{lll|rrr}
        Write   & Write  & Read    &  &  & \\
        target  & mode   & target  & E2E lat (s)    & Per-req. lat (ms) & \# inconsist. \\ \midrule[1pt]
        Storage & Sync   & Storage & 74.4 $\pm$ 0.2 & 80.0 $\pm$ 29.5   & 0 $\pm$ 0   \\
        Owner   & Sync   & Owner   & 42.6 $\pm$ 0.2 & 41.6 $\pm$ 12.8   & 0 $\pm$ 0   \\
        Owner   & Async  & Owner   & 39.7 $\pm$ 0.6 & 38.5 $\pm$ 11.4   & 1.3 $\pm$ 1.9 \\
        Owner   & Sync   & Local   & 32.7 $\pm$ 0.3 & 32.7 $\pm$ 9.0    & 800 $\pm$ 0   \\
        Local   & Sync   & Local   & 31.1 $\pm$ 0.3 & 30.3 $\pm$ 13.1   & 800 $\pm$ 0   \\
        Local   & Aync   & Local   & 31.0 $\pm$ 0.3 & 30.3 $\pm$ 13.2   & 800 $\pm$ 0   \\ \bottomrule
    \end{tabular}
    } %
    \caption{
      Latency (end-to-end and per-request) and number of inconsistencies for different write/read settings for a JupyterLess notebook counting to 1K with five instances sharing state.
      Inconsistencies are the absolute difference between the final counter value and 1K.
      Performance increases as consistency and fault tolerance decrease.
    }
    \label{tab:consistency}
    \vspace{-.15in}
\end{table}

\vspace{-.1in}
\subsection{Trading off consistency and performance} \label{sec:consistencyeval}
\faast allows users to trade off consistency for performance.
We evaluate the write/read settings from Table~\ref{tab:consistency-versions} using a JupyterLess notebook in which five cachelets share the state for counting up to 1K.
Application instances add to the counter in round-robin fashion, and we expect the counter at the end to have a value of 1K.
We measure inconsistencies as the absolute difference between the final counter value and 1K.
This is a critical primitive in multiplayer games~\cite{donkervliet2020minecraft}.

Table~\ref{tab:consistency} shows the end-to-end latency, per-request latency, and number of inconsistencies for all five settings.
As expected, latency drops as we relax consistency requirements.  
For example, writing to the local cache and reading from the local cache is fastest, but provides no consistency (800 inconsistencies recorded) and the lowest fault tolerance.
Writing to the local cache asynchronously and reading from the local cache is equivalent to Cloudburst's performance. Cloudburst would exhibit better consistency due to its lattice datatypes, but requires support from the datastore.
Latency varies the most when writing and reading from remote storage, and the least when writing and reading from the local cache.

\begin{figure}
  \centering
  \begin{subfigure}[t]{0.47\linewidth}
    \includegraphics[width=1\columnwidth]{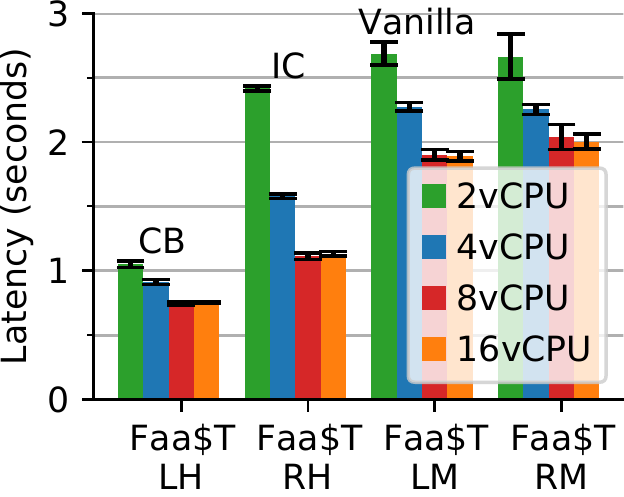}
    \caption{AMBER Alert pipeline}
    \label{fig:instsweep-mlinf}
  \end{subfigure}
  \begin{subfigure}[t]{0.47\linewidth}
    \includegraphics[width=1\columnwidth]{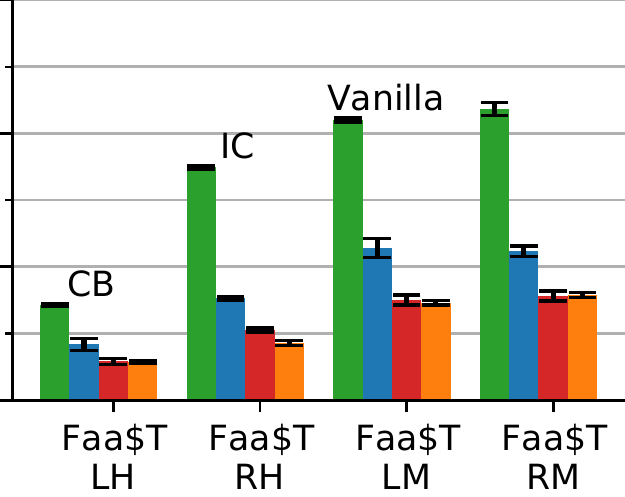}
    \caption{350MB DataFrame Notebook}
    \label{fig:instsweep-parquet}
  \end{subfigure}
  \caption{
    Latency as the instance size varies for (a) the AMBER Alert pipeline, and (b) summing a 350MB DataFrame column in a JupyterLess notebook.
    Instances memory and network bandwidth scale linearly as the number of cores increases.
    \faast benefits from instances with higher bandwidth.
  }
  \label{fig:instsweep-all}
  \vspace{-.15in}
\end{figure}

\vspace{-.1in}
\subsection{Sensitivity to instance size} \label{sec:insttypeseval}
Finally, we evaluate the sensitivity of running applications with \faast as the instance size varies.
We run the AMBER Alert pipeline and sum a 350MB DataFrame column in a JupyterLess notebook.
Instances scale linearly in terms of memory and network bandwidth as the vCPUs increase.
For example, the 2vCPU instance has 8GiB of memory and 1Gbps of network bandwidth, and the 4vCPU instance has 16GiB of memory and 2Gbps of network bandwidth.

Figures~\ref{fig:instsweep-mlinf} and~\ref{fig:instsweep-parquet} show that as instances increase in size, \faast's latency decreases for both applications.
Larger instances have higher network bandwidth, which is beneficial for data accesses to remote storage and between instances.
Data accesses of these two applications saturate the bandwidth with the 8vCPU instance.
Thus, although the 16vCPU instance is the highest bandwidth instance size, the performance remains the same as the 8vCPU.
Some cloud providers offer instances as small as 2vCPU with up to 10Gbps bandwidth, allowing \faast to have high performance even on small instances suitable for FaaS.

\section{Related Work}
\myparagraph{Ephemeral serverless storage}
In Section~\ref{sec:existingsys} we describe the limitations of several existing storage and data caching solutions for FaaS~\cite{klimovic2018pocket,wang2020infinicache,pu2019locus,shillaker2020faasm,sreekanti2020cloudburst}.
Unlike these systems, \faast does not require external resources beyond what is provided to the invoked function, is transparent to applications, and can scale as the data size and access patterns vary.

OFC\footnote{This work was done concurrently and independently of OFC.} is the closest work to \faast.
It transparently caches objects using RAMCloud~\cite{ousterhout2015ramcloud} and leverages machine learning to dynamically size the cache.
Unlike OFC, \faast pre-warms objects when an application is loaded, supports large ($>$ 10MB) object caching and optimizes for their data transfer latency from remote storage with bandwidth scaling, and only needs to keep one copy of data in shared memory (compared to OFC that requires a copy in the worker and in RAMCloud).
\faast also incurs lower decision overheads and is easier to manage by not requiring the use of machine learning for its decision-making.

\myparagraph{Serverless frameworks}
Several frameworks have recently emerged enabling users to run, for example,
linear algebra~\cite{jonas2017pywren}, video
encoding~\cite{fouladi2017excamera}, video analytics
pipelines~\cite{ao2018sprocket}, ML training~\cite{carreira2019cirrus}, and
general burst-parallel applications~\cite{fouladi2019gg} on up to thousands
of serverless functions.
These, and their applications, would benefit from managing and transferring intermediate data between serverless functions using \faast.
Since \faast is transparent to applications, little to no changes would be needed to interact with \faast.

\myparagraph{Improving serverless performance}
There have been many approaches to reduce the execution time of serverless function, such as making containers more lightweight~\cite{agache2020firecracker,oakes2018sock}, using snapshotting techniques~\cite{cadden2020seuss,du2020catalyzer}, or reducing the number of cold-starts~\cite{shahrad2020serverless,fuerst2021faascache}.
Shredder~\cite{zhang2019shredder} focuses on how to provide isolation for multi-tenancy.
Lambada~\cite{muller2020lambada} focuses on improving performance for serverless applications with exchange operators.
Lambdata~\cite{tang2020lambdata} allows users to expose their data read and write intents for making optimizations such as co-locating functions working on the same data.
These optimizations are orthogonal to \faast, which focuses on how to improve state management for serverless functions, and how to scale instances to improve application performance.

\myparagraph{Consistency and fault tolerance}
Consistency and fault tolerance protocols have been heavily studied.
Recent work has explored how to enable both of these for serverless applications.
Faasm~\cite{shillaker2020faasm} and Cloudburst~\cite{sreekanti2020cloudburst} provide local caches backed by a distributed key-value store. Faasm allows for strong consistency by using global locks at the KVS; Cloudburst provides guarantees for repeatable reads and causal consistency by using lattice data types supported by its local caches and by its Anna KVS backend~\cite{wu2018anna}.
AFT~\cite{sreekanti2020aft} added a fault tolerance shim layer for FaaS, and implemented protocols for read atomic isolation.
Beldi~\cite{zhang2020beldi} provides a framework to write transactional and fault tolerant stateful serverless functions by extending Olive~\cite{setty2016olive} with a novel data structure to support fast logging and exactly-one semantics.
\faast transparently supports different consistency and fault tolerance settings directly in the functions runtime, and can be extended to support future protocols.

\section{Conclusion}
We presented \faast, a transparent caching layer for serverless applications.
We motivated its design with a characterization of production applications.
We tie \faast to the application, scale based on compute demands and data access patterns, and provide data consistency that can be set per application.
We implemented it in a production serverless platform.  Compared to existing systems, \faast is on average 57\% faster and 99.99\% cheaper when running two challenging applications.

\bibliographystyle{abbrv} %
\bibliography{paper}

\end{document}